\documentclass[a4paper,11pt]{article}

\usepackage{jheppub}
\usepackage{graphicx}
\usepackage{xcolor}
\usepackage{braket}
\usepackage{subfigure}

\usepackage[export]{adjustbox}

\graphicspath{{graphs/}}

\providecommand{\nn}{\nonumber}  % not numbering lines in the align environment

\newcommand{\tvec}[1]{\boldsymbol{#1}}

\newcommand{\deltaf}{\delta^{(D)}}      % delta^(D)

\newcommand{\dd}{\mathrm{d}}     % differential "d"

\newcommand{\der}[2]{\frac{\dd #1}{\dd #2}}
\newcommand{\ddf}{\dd^D}  % differential "d^4"

\newcommand{\twopif}{(2 \pi)^D}    % (2pi)^D

\newcommand{\fourmomdiff}[1]{\frac{\ddf #1}{\twopif}}
\newcommand{\ms}{\mskip 1.5mu}
\newcommand{\bs}{\mskip -1.5mu}

\newcommand{\rev}[1]{#1}

\preprint{DESY 18-204}

\title{Factorisation of soft gluons in multiparton scattering}

\author[a]{Markus Diehl}
\author[a]{and Riccardo Nagar}

\affiliation[a]{Deutsches Elektronen-Synchrotron DESY,\\
Notkestra{\ss}e 85, 22607, Hamburg, Germany}

\emailAdd{markus.diehl@desy.de}
\emailAdd{riccardo.nagar@desy.de}

\abstract{We show that soft gluons exchanged between the two colliding protons in  multiparton scattering processes can be decoupled, such that their effects are described by the vacuum expectation values of Wilson lines.  Our argument relies on nonabelian Ward identities and generalises the proof of factorisation for single Drell-Yan production given by Collins, Soper and Sterman.}

\begin{document}
\maketitle
\flushbottom

\section{Introduction}

Significant progress has been made in understanding double parton scattering (DPS), the process in which two pairs of partons take part in two hard scattering processes in a single proton-proton collision.  Experimental information comes from an increasing number of DPS studies at the LHC, whilst on the theory side work is being done both on conceptual aspects and on phenomenology and modelling.  The current status of the field is documented in the monograph \cite{Bartalini:2017jkk} and in the talks at the most recent Workshop on Multiple Partonic Interactions at the LHC (MPI@LHC 2018\footnote{\url{https://indico.cern.ch/event/736470}}).

The theoretical description of DPS relies on factorisation formulae akin to those for single parton scattering (SPS).  For one of the simplest SPS processes, namely  Drell-Yan production, detailed proofs of factorisation are available (see the original work \cite{Collins:1985ue,Collins:1988ig,Bodwin:1984hc} and the review in \cite{Collins:2011zzd}), and it is natural to ask to which extent they can be generalised to DPS.  The work in \cite{Diehl:2011yj,Diehl:2015bca,Diehl:2017kgu,Vladimirov:2017ksc,Buffing:2017mqm} shows that indeed factorisation for the double Drell-Yan process can be established at a comparable level of detail as for single Drell-Yan.  An important aspect left open in that work is however a proof that the effects of soft gluon exchange between the spectator partons in the two protons can be expressed in a so-called soft factor, which is defined as the vacuum expectation value of Wilson line operators.  We refer to this as the decoupling of soft gluons.  Given that DPS has a considerably more complicated colour structure than SPS, a corresponding proof is highly desirable.  Indeed, a failure to decouple soft gluons from the spectator partons would imply that the two colliding protons are tied together by nonperturbative dynamics, which would seriously curtail the predictive power of theory.

To close this gap and show that soft gluons can be decoupled in DPS processes is the aim of the present work.  We make use of nonabelian Ward identities, following closely the derivation of soft gluon decoupling in the work by Collins, Soper and Sterman \cite{Collins:1985ue,Collins:1988ig} with some important differences pointed out later.  We note that a powerful alternative method  is given by soft-collinear effective theory (SCET), where the decoupling of soft gluons is achieved by field redefinitions \cite{Bauer:2002nz}.  A general formulation of DPS factorisation in SCET is given in \cite{Manohar:2012jr}.  The systematic treatment of the so-called Glauber momentum region in this framework has however only been formulated recently \cite{Rothstein:2016bsq}, and a corresponding analysis for single or double Drell-Yan is still missing.  Using conventional methods, the role of the Glauber region for DPS was analysed in \cite{Diehl:2015bca}.

This work is structured as follows.  In section~\ref{sec:overview} we give an overview of the factorisation proof for double Drell-Yan production, thus putting the analysis of the present paper into a larger context.  In section~\ref{sec:collinear-wilson} we introduce our notation for Wilson and eikonal lines, and we give the expression of the DPS amplitude after longitudinally polarised collinear gluons have been decoupled from the hard scattering.  Section~\ref{sec:proof} shows how soft gluons in that expression can be decoupled from collinear partons, which is the main result of our paper.  In section~\ref{sec:final_result} we show how the momentum-space eikonal lines obtained in the previous step can be converted into Wilson lines.  We briefly summarise our work in section~\ref{sec:sum}.  In an appendix we derive Ward identities for correlation functions that involve Wilson lines, which is required for the main proof in section~\ref{sec:proof}.

\section{Overview of factorisation proof}
\label{sec:overview}

The decoupling of soft gluons is one step in the factorisation proof for the double Drell-Yan process.  To put this into context, we briefly sketch the different steps of the overall proof.  We largely follow the presentation in section~2.1 of \cite{Diehl:2015bca}, which builds on the factorisation proof for single Drell-Yan production discussed in chapter~14.4 of \cite{Collins:2011zzd}.

We consider both collinear factorisation, where the transverse momenta of the Drell-Yan lepton pairs are integrated over, and TMD factorisation, where they are measured and small compared with the lepton pair invariant masses.  The factorisation proof is readily extended to other channels in which colourless particles are produced.  Examples of hard scattering processes are $q\bar{q} \to Z$, $q\bar{q}\ms' \to W$, $g g\to H$ and $g g\to \gamma\gamma$.

\begin{enumerate}
\item \textbf{Power counting.}  The approximations leading to hard scattering factorisation are based on an expansion in powers of $\Lambda /Q$.  Here $\Lambda$ represents all small scales, i.e.\ hadron masses and the scale of non-perturbative interactions.  In TMD factorisation this includes the measured transverse momenta $\tvec{q}_1$ and $\tvec{q}_2$ of the produced bosons.  The large scale $Q$ stands for the invariant masses $Q_1$ and $Q_2$ of the two produced systems.  In the present work, we will not keep track of the specific power of $\Lambda /Q$ by which neglected terms are suppressed.
\item \textbf{Leading graphs and momentum regions.}  Following the analysis of Libby and Sterman \cite{Libby:1978bx,Libby:1978nr}, one identifies the leading graphs contributing to the process and the leading momentum regions in each leading graph.  For each leading momentum region, the graph can then be decomposed into subgraphs associated with hard, collinear or soft momenta.  Taking a frame in which the proton with momentum $P$ moves along $z$ and the proton with momentum $\bar{P}$ along $-z$, these momenta are characterised as follows:
\begin{align}
\label{mom-regions}
   \text{hard}             &\qquad |\ell^+| \sim |\ell^-|
   \sim |\tvec{\ell}| \sim Q \,,
\nn \\
   \text{right-collinear}  &\qquad |\ell^+| \sim Q \,, \quad
   |\ell^-|\ll Q \,,  \quad \tvec{\ell}^2 \sim |\ell^+ \ell^-| \,,
\nn \\
   \text{left-collinear}   &\qquad |\ell^+| \ll Q \,, \quad
   |\ell^-|\sim Q \,, \quad \tvec{\ell}^2 \sim |\ell^+ \ell^-| \,,
\nn \\
   \text{central soft}     &\qquad |\ell^+| \sim |\ell^-|
   \sim |\tvec{\ell}| \ll Q \,,
\nn \\
   \text{Glauber soft}     &\qquad |\ell^+ \ell^-| \ll \tvec{\ell}^2 \,,
   \quad |\tvec{\ell}| \ll Q \,.
\end{align}
Here we use light-cone coordinates $v^{\pm} = (v^0 \pm v^3) /\sqrt{2}$ for each four-vector $v^\mu$ and write its transverse part as $\tvec{v} = (v^1, v^2)$.  Among soft momenta, the Glauber region plays a special role, which will be recalled shortly.  In  \eqref{mom-regions} we do not specify the scaling of small momentum components in terms of $\Lambda$ and $Q$, referring to \cite{Diehl:2015bca} and to \cite{Collins:2011zzd} for detailed discussions. \rev{For the arguments in the present work, the specification in \eqref{mom-regions} is sufficient.   We note that the scaling of momenta plays an essential role in the analysis of collinear or TMD factorisation in SCET, see e.g.\ \cite{Bauer:2002nz,Becher:2010tm,Chiu:2012ir}.}

Important features of leading momentum configurations are that soft momenta may not enter a hard subgraph, and that the only  partons that can enter a soft subgraph are gluons.  Exactly one quark and one antiquark enters each hard subgraph in the amplitude for Drell-Yan production; for other channels like Higgs production one has two transversely polarised gluons instead.  Additional gluons may connect a collinear with a hard subgraph.

\begin{figure}
\centering
\includegraphics[scale=0.5]{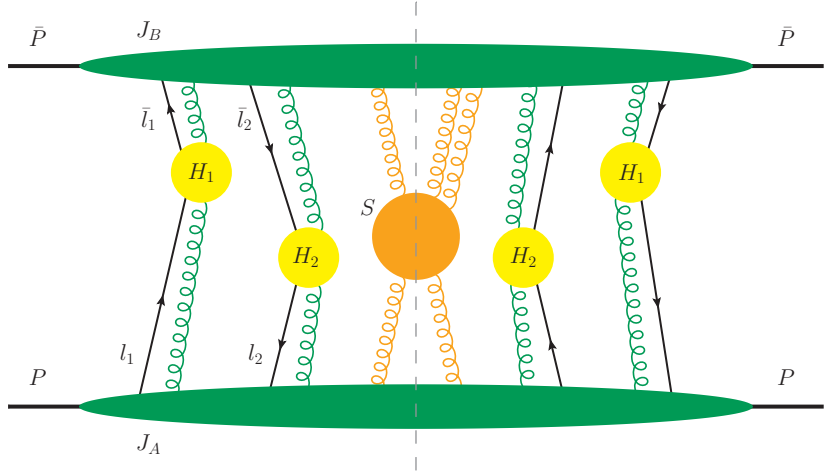}
\caption{\label{fig:leading_regions} Structure of the leading graphs in TMD factorisation for the double Drell-Yan process.  $H_1$ and $H_2$ denote the combination of hard subgraphs to the left and to the right of the final-state cut, which is represented by the vertical dashed line.  \protect\rev{For the sake of clarity, the electroweak gauge bosons emerging from $H_1$ and $H_2$ are not shown here and in the next figures.}}
\end{figure}

The structure of the leading graphs for the double Drell-Yan process in TMD factorisation is shown in figure~\ref{fig:leading_regions}.  \rev{We note that the soft subgraph $S$ connecting $J_A$ and $J_B$ may consist of several connected parts, but that each of these must connect $J_A$ with $J_B$.  In collinear factorisation, the leading graphs are more complicated.  The final state can include unobserved jets produced in one of the hard-scattering processes, and soft gluons can connect the collinear graphs associated with these jets to the collinear graphs $J_A$ and $J_B$ of the incoming protons, as shown for instance in figure~2 of \cite{Diehl:2015bca}.  A series of arguments is needed to decouple the soft gluons from the unobserved jets and to remove the associated collinear subgraphs.  These arguments, which in particular make use of unitarity, proceed in the same manner for single and double Drell-Yan production (and are very similar to those for inclusive $e^+ e^-$ annihilation into hadrons).  As our focus is on double parton scattering, we shall not expand on this issue here.  The leading graphs at the end of this procedure have the same form as in TMD factorisation, except that the hard scattering subgraphs extend over the final state cut, as shown in figure~3 of \cite{Diehl:2015bca}.}

The final state of two heavy bosons plus unspecified hadrons can not only be produced by double parton scattering (DPS) but also by the conventional single parton scattering (SPS) mechanism.  There are Feynman graphs that have leading momentum regions both of DPS and of SPS type, which leads to a double counting issue.  A scheme to handle this issue is presented in \cite{Diehl:2017kgu}.  Since the decoupling of soft gluons is not affected by this, we shall not go into further detail here.
\item \label{approx} \textbf{Kinematic approximations.}  For each momentum region, a set of kinematic approximations is applied to the graph.  Corrections to these approximations are power suppressed.  Collinear lines entering a hard subgraph are approximated by
\begin{equation}
   \ell \to (\ell^+, 0, \boldsymbol{0})
\end{equation}
for right-collinear momenta and
\begin{equation}
   \bar{\ell} \to (0, \ell^-, \boldsymbol{0})
\end{equation}
for left-collinear ones, where we write the components of a four-vector as $(v^+, v^-, \tvec{v})$.  For soft lines entering a collinear subgraph, we approximate
\begin{align}
\label{eqn:soft_approximation_momenta}
   \ell &\to \tilde{\ell} = (0, \ell^-, \boldsymbol{\ell}),
\nn \\
   \bar{\ell} &\to \tilde{\bar{\ell}} = (\ell^+, 0, \boldsymbol{\ell}),
\end{align}
where the first line refers to the soft gluons entering the right-collinear graph $J_A$ and the second line to the ones entering the left-collinear graph $J_B$.
It is important to note that the approximation in \eqref{eqn:soft_approximation_momenta} is weaker than the one in the work by Collins, Soper and Sterman \cite{Collins:1985ue,Collins:1988ig,Collins:2011zzd}, who set the transverse momentum $\tvec{\ell}$ on the r.h.s.\ of \eqref{eqn:soft_approximation_momenta} to zero.  \rev{The justification for this stronger approximation relies on a specific routing of loop momenta, for which the collinear subgraphs are independent of $\tvec{\ell}$ after soft gluons have been decoupled.  Such a routing is possible for SPS but not for DPS, as shown for a simple example in section~2.3 of \cite{Diehl:2015bca}. The calculation in section~\ref{sec:final_result} shows that keeping track of soft transverse momenta in the collinear factors is essential for getting the correct position arguments of Wilson lines in the soft factors.  One goal of the present work is to establish that the weaker approximation in \eqref{eqn:soft_approximation_momenta} is sufficient for showing the decoupling of soft gluons.}

In practice, loop integrations are performed over the full phase space after the above approximations have been made, in order to avoid the use of explicit cutoffs.  The contribution from the ``unwanted'' momentum regions is removed by a recursive subtraction mechanism, as discussed in detail in \cite{Collins:2011zzd} and briefly explained in \cite{Diehl:2015bca}.  These subtractions are also relevant when subgraphs limited to a certain momentum region are expressed in terms of operator matrix elements, because the latter imply integration over the full region of all internal momenta.  We return to this issue in point~\ref{mat-el-defs} below.
\item \textbf{Dirac and Lorentz indices.}  Fierz transformations are applied to the fermion lines connecting collinear with hard subgraphs, and approximations are then made regarding the leading Lorentz components of the Dirac matrices $\gamma^\mu$ arising from these transformations.  Corresponding steps are taken for the Lorentz indices of gluon fields if the hard scattering is initiated by gluons.  Details are given in section~2.2.1 of \cite{Diehl:2011yj} for double parton scattering.  In the present context, they are not important.
\item \label{contour-def} \textbf{Glauber region.}  To enable step \ref{decouple-soft} below, the loop integrations over plus or minus components of soft momenta are deformed in the complex plane so as to avoid the Glauber region.  To establish that such a deformation is possible requires a sum over all final state cuts of a given graph and is a crucial step in any factorisation proof.  Detailed analyses  are given in \cite{Collins:1985ue,Collins:1988ig,Collins:2011zzd} and \cite{Bodwin:1984hc} for the single and in \cite{Diehl:2015bca} for the double Drell-Yan process.  Into which region the Glauber momenta are to be deformed is subtle, as was pointed out in section~4.1 of \cite{Boer:2017hqr}.  This issue concerns both single and double hard scattering and merits further investigation, which shall not be attempted here.

The discussion of complex contour deformation in the papers just cited relies on the use of Feynman gauge for the gluon field, which we therefore use in the present work as well.
\item \label{decouple-coll} \textbf{Collinear decoupling.}  Collinear gluons are decoupled from the hard subgraphs, using so-called Grammer-Yennie approximations \cite{Grammer:1973db} and nonabelian Ward identities.  This is discussed for single Drell-Yan production in the appendix of \cite{Collins:1985ue}.  The argument can be immediately transferred to the double Drell-Yan process since the Ward identity is applied to a hard scattering graph, which can be done sequentially for the two hard scatters in DPS.  After this procedure, the collinear subgraphs are modified and contain eikonal lines.  The leading graphs for the double Drell-Yan cross section then have the form shown in figure~\ref{fig:cross_section_collinear}.  More detail is given in section~\ref{sec:collinear}.

\begin{figure}
\centering
\includegraphics[scale=0.5]{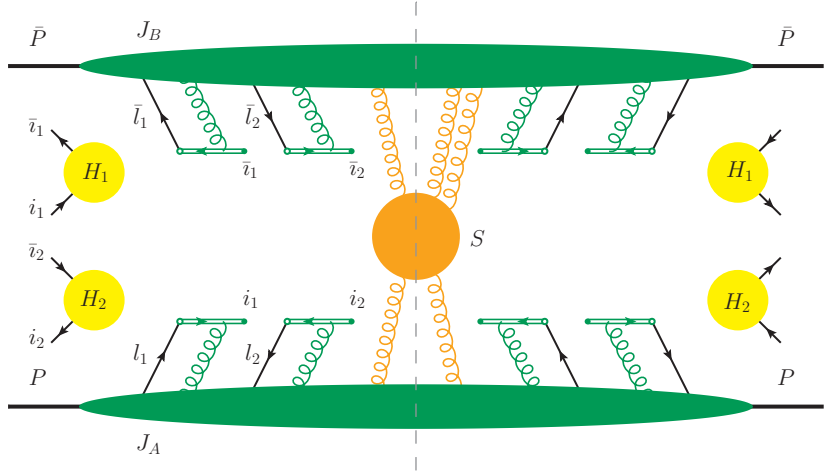}
\caption{\label{fig:cross_section_collinear} Structure of the TMD cross section after collinear gluons have been decoupled from the hard subgraphs.}
\end{figure}
\item \label{decouple-soft} \textbf{Soft decoupling.}  Soft gluons are decoupled from the collinear subgraphs, again using Grammer-Yennie approximations and nonabelian Ward identities.  In this case, the latter are applied to the collinear graphs and are therefore more involved for DPS than for SPS processes.  After this step, which is the subject of the present work, the gluons emerging from the soft subgraph couple to eikonal lines as shown in figure~\ref{fig:cross_section_factorised}.  Both collinear and soft subgraphs are now decoupled from each other.

Let us briefly review the Grammer-Yennie approximation for a soft gluon with momentum $\ell$ flowing from the soft into the right-collinear subgraph.  We have
\begin{align}
\label{soft-GY}
   S_\mu (\ell) \, J_{A}^{\mu} (\tilde{\ell})
& \approx S^- (\ell)\, J_A^+ (\tilde{\ell})
 = S^- \rev{(\ell)} \, \frac{v_{R}^{+} \,
   \tilde{\ell}^-}{v_R^+ \, \ell^-_{} + i \epsilon} \, J_{A}^{+} (\tilde{\ell})
\nonumber \\
& \approx S_\mu \rev{(\ell)} \, \frac{v_{R}^{\mu} \,
  \tilde{\ell}_\nu}{v_R \cdot \ell + i \epsilon} \,
  J_{A}^{\ms \nu} (\tilde{\ell}) \,,
\end{align}
where $\tilde{\ell}$ is the approximation \eqref{eqn:soft_approximation_momenta} of the gluon momentum in collinear graphs, and $v_R^{} = (v_R^+, v_R^-, 0, 0)$ is a spacelike auxiliary vector satisfying $v_R^+  \gg - v_R^- > 0$.  In  \eqref{soft-GY} we have in particular used that the components of $J_A^\mu$ scale like a collinear momentum.  The last step can then be justified for $|\ell^-| \sim |\tvec{\ell}|$ but is not generally valid in the Glauber region, where one can have $|\ell^-| \ll |\tvec{\ell}|$.  \rev{This is the reason for first making the complex contour deformation mentioned in step~\ref{contour-def}.  To avoid propagator poles in both $J_{A}$ and $J_{B}$, this deformation is into the upper half plane for $\ell^-$ and the lower half plane for $\ell^+$.  The signs of $i \epsilon$ and of $v_R^+$ and $v_R^-$ in \eqref{soft-GY} are chosen such that one can deform $\ell^-$ and $\ell^+$ back to the real axis without obstruction (see step~\ref{deform-back}).}

To simplify the notation in later sections, we implement the Grammer-Yennie approximation by replacing the metric tensor in the Feynman gauge gluon propagator as
\begin{align} \label{eqn:grammer_yennie_def}
g^{\mu\nu} \to Y_R^{\mu\nu} (\ell)
 & = \frac{v_R^\mu \, \tilde{\ell}^{\ms \nu}_{}}{v_R \cdot \ell + i \epsilon} \,,
&
g^{\mu\nu} \to Y_L^{\mu\nu} (\bar{\ell})
 & = \frac{v_L^\mu \, \tilde{\bar{\ell}}^{\ms \nu}_{}}{v_L \cdot \bar{\ell}
     \rule{0pt}{0.95em} + i \epsilon}
\end{align}
for right and left collinear gluons, respectively, where $v_L^{} = (v_L^+, v_L^-, 0, 0)$ is a spacelike auxiliary vector with $v_L^-  \gg - v_L^+ > 0$.  In both cases, $\mu$ is to be contracted with the soft and $\nu$ with the collinear subgraph, and the soft momentum is routed to flow out of~$S$.

\begin{figure}
\centering
\includegraphics[scale=0.5]{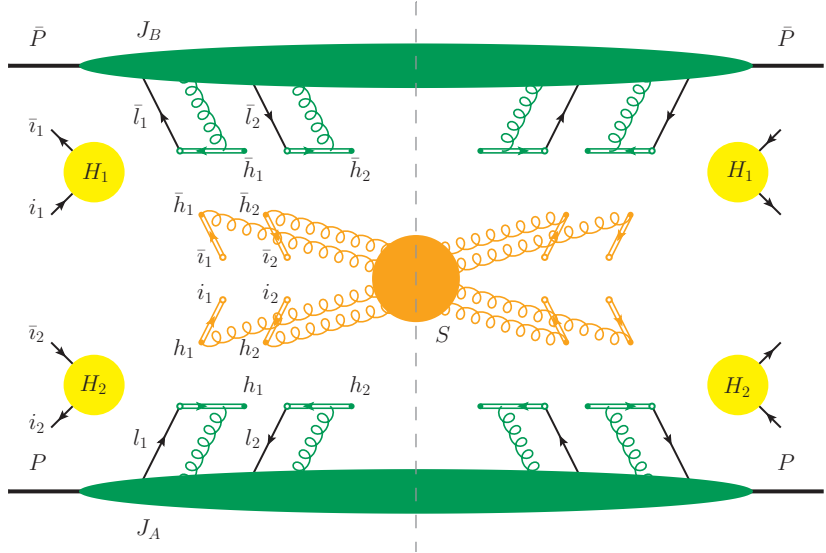}
\caption{\label{fig:cross_section_factorised} Structure of the TMD cross section after soft gluons have been decoupled from the collinear subgraphs.}
\end{figure}
\item \label{deform-back} \textbf{Back to real momenta.}  At this point, the loop integrations deformed into the complex plane in step~\ref{contour-def} can be brought back to the real axis.  As we will see in section~\ref{sec:proof}, the application of Grammer-Yennie approximations and Ward identities in step~\ref{decouple-soft} is intertwined, so that the deformation back to the real axis cannot be done before.\footnote{This was overlooked in section 2.1 of \protect\cite{Diehl:2015bca} but does not affect the remaining arguments in that work.}
\item \label{mat-el-defs} \textbf{Matrix elements.}  The collinear and soft subgraphs can be expressed in terms of operator matrix elements.  The eikonal lines in these subgraphs (see figure~\ref{fig:cross_section_factorised}) are generated by Wilson line operators.  This step provides a non-perturbative definition of the soft and collinear factors, which can indeed not be computed in perturbation theory.

When representing the collinear factors by matrix elements, one needs to subtract the contribution from regions in which gluons are soft rather than collinear.  This is handled by the subtraction formalism mentioned above.  The same arguments as in step~\ref{decouple-soft} can be used to simplify the subtraction terms and express them again in terms of operator matrix elements.  This was already sketched in section 2.2 of \cite{Diehl:2015bca} and will be briefly reviewed in section~\ref{sec:soft-gluon-sub} here.
\item \textbf{Rapidity dependence.}  The introduction of eikonal lines in earlier steps generates so-called rapidity divergences if it is done in a naive way.  These divergences need to be regularised.  Following the work in \cite{Diehl:2015bca}, we use the regulator of Collins~\cite{Collins:2011zzd} and take non-lightlike vectors in the eikonal lines, as already mentioned above.  Other rapidity regulators have been used in the literature, see e.g.\ \cite{Chiu:2011qc,Chiu:2012ir,Echevarria:2015byo,Echevarria:2016scs} for SPS and \cite{Manohar:2012jr,Vladimirov:2016qkd,Vladimirov:2017ksc} for DPS.

By a nontrivial rearrangement, the soft factor can be split into pieces and absorbed into the collinear factors.  The rapidity regulator can then be removed, but a dependence on a parameter related to rapidities (often called $\zeta$) remains in the individual factors.  This dependence is closely related with the resummation of Sudakov logarithms and hence has physical significance.  A construction for DPS was worked out in \cite{Buffing:2017mqm}, generalising the formalism of Collins \cite{Collins:2011zzd} for single Drell-Yan production.   \rev{Interestingly, DPS cross sections contain Sudakov logarithms not only in TMD but also in collinear factorisation, in contrast to SPS.  This was already pointed out in \cite{Mekhfi:1988kj,Manohar:2012jr} and is due to the more complicated colour structure of DPS.}
\item \textbf{Ultraviolet renormalisation.}  Up to this point, all subgraphs contain ultraviolet divergences, which are now renormalised following standard procedures.  This can be done using dimensional regularisation and $\overline{\text{MS}}$ subtraction.  \rev{As is well known from SPS, the details of renormalisation differ between TMD and collinear factorisation.  In both cases,} the result is a factorisation formula for the cross section, in which double parton distributions contain soft and collinear dynamics, whereas hard scattering cross sections describe the dynamics associated with hard momenta.
\end{enumerate}

This concludes our broad overview of factorisation for double parton scattering.  In the next section, we recall some details about eikonal and Wilson lines and set up our notation.  After that, we show the decoupling of soft gluons, which is the main result of this work.

\section{Wilson lines and decoupling of collinear gluons}
\label{sec:collinear-wilson}

\subsection{Wilson and eikonal lines}
\label{sec:wilson}

\begin{figure}
   \centering
   \includegraphics[width=0.99\textwidth]{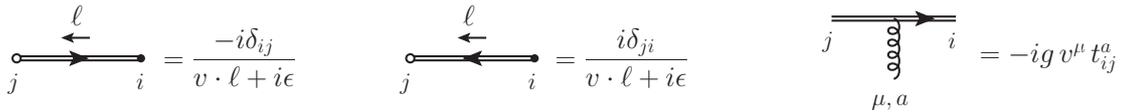}
   \caption{Feynman rules for eikonal lines in the fundamental representation of the colour group.}
   \label{fig:eikonal_rules}
\end{figure}

The decoupling of collinear and of soft gluons using Grammer-Yennie approximations and Ward identities leads to subgraphs in which collinear or soft gluons couple to eikonal lines.  The Feynman rules for such lines in a scattering amplitude are given in figure~\ref{fig:eikonal_rules}, where we follow the conventions spelled out in \cite{Diehl:2011yj,Buffing:2017mqm}: the arrow on the line indicates colour flow, whereas the open and closed dots at the end specify the sign of $i\epsilon$.  One readily sees the correspondence between these Feynman rules and the Grammer-Yennie factor in \eqref{eqn:grammer_yennie_def}.

When collinear or soft subgraphs are represented as operator matrix elements, the eikonal lines turn into Wilson line operators after summing over all orders of the strong coupling $g$.  Wilson lines in the fundamental colour representation are defined by
\begin{align} \label{eqn:wilson_line_def}
   W_{v,\, i j} (x) &= \mathcal{P} \exp \left[ i g \!
   \int_{-\infty}^0 \! \dd s \, v \cdot A^a (x + s v) \, t^a_{i j} \right],
\end{align}
where the matrices $t^a$ are the generators of the colour group and $\mathcal{P}$ denotes path-ordering of colour matrices.  Wilson lines $W_{v}^{b c} (x)$ in the adjoint representation are defined in analogy, with $t^a_{i j}$ in \eqref{eqn:wilson_line_def} being replaced by $-i f^{a b c}$.  The direction of the path in \eqref{eqn:wilson_line_def} is appropriate for Drell-Yan production and results from the sign of $i\epsilon$ in the Grammer-Yennie approximation \eqref{eqn:grammer_yennie_def}.

\begin{figure}
   \centering
   \includegraphics[scale=0.6]{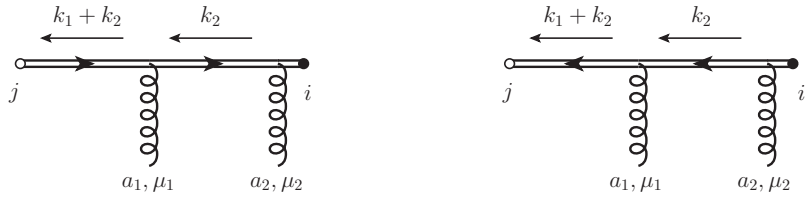}
\caption{Graphical representation of eikonal lines ${\mathcal{E}^{(2)}_{i j}} (k_1, k_2)$ and $\bar{\mathcal{E}}^{(2)}_{j i} (k_1, k_2)$ with two gluon insertions, defined by \protect\eqref{eqn:eikonal_order_n} and \protect\eqref{eqn:adj_eikonal_order_n}.}
\label{fig:eikonal_order_2}
\end{figure}

To make the correspondence between Wilson and eikonal lines explicit, we expand the former in the coupling,
\begin{align}
\label{WL-expand}
W_{v,\, i j}^{} (x) &= \sum_{n=0}^\infty \, g^n_{} \,
   W_{v,\, i j}^{(n)} (x) \,,
\end{align}
and perform a Fourier transform to momentum space.  For a general operator $\phi(x)$ in position space, we define the momentum-space version
\begin{align}
\label{fourier-def}
\phi(k) &= \int \ddf{z} \; e^{i k \cdot x} \, \phi(x) \,.
\end{align}
The lowest-order term of the Fourier transformed Wilson line is trivial and reads $W_{v,\, i j}^{(0)}(k) = \delta_{i j} \,\twopif \, \deltaf(k)$, whilst for $n > 0$ we have
\begin{align} \label{WL-FT}
W_{v,\, i j}^{(n)} (k)
&=  \int \! \fourmomdiff{k_1} \dots \fourmomdiff{k_n} \, \twopif \,
   \deltaf(k - \sum k_i)
\nn \\[0.2em]
&\quad \times
   \Bigl[ \mathcal{E}^{(n)}_{v,\, i j} \bigl(\{ k_n \}\bigr)
   \Bigr]^{a_1 \dots a_n}_{\mu_1 \dots \mu_n}
   \, A^{a_1 \mu_1} (k_1) \dots A^{a_n \mu_n} (k_n) \,,
\end{align}
where
\begin{align}
\label{eqn:eikonal_order_n}
      \Bigl[ {\mathcal{E}^{(n)}_{v,\, i j}} \bigl(\{ k_n \}\bigr)
      \Bigr]^{a_1 \cdots a_n}_{\mu_1 \cdots \mu_n}
&= \frac{- v_{\mu_n}}{k_n \cdot v + i \epsilon} \,
   \frac{- v_{\mu_{n-1}}}{(k_n + k_{n-1}) \cdot v + i \epsilon}
   \cdots
   \frac{- v_{\mu_1}}{\rev{(k_n + \cdots + k_1)} \cdot v + i \epsilon}
\nn \\[0.1em]
& \quad \times
    \bigl( t^{a_n} \cdots t^{a_1} \bigr)_{i j}
\end{align}
denotes an eikonal line with $n$ gluon insertions.  Here and in the following, curly brackets in function arguments denote a set of momenta, i.e.\ $\{ k_n \}$ stands for $k_1, \ldots, k_n$.  For conjugate Wilson lines, we have
\begin{align} \label{WL-adj-FT}
W_{v,\, j i}^{\dagger (n)} (k)
&=  \int \! \fourmomdiff{k_1} \dots \fourmomdiff{k_n} \, \twopif \,
   \deltaf(k - \sum k_i)
\nn \\[0.2em]
&\quad \times
   \Bigl[ \bar{\mathcal{E}}^{(n)}_{v,\, j i} \bigl(\{ k_n \}\bigr)
   \Bigr]^{a_1 \dots a_n}_{\mu_1 \dots \mu_n}
   \, A^{a_1 \mu_1} (k_1) \dots A^{a_n \mu_n} (k_n)
\end{align}
with
\begin{align}
\label{eqn:adj_eikonal_order_n}
      \Bigl[ \bar{\mathcal{E}}^{(n)}_{v,\, j i} \bigl(\{ k_n \}\bigr)
      \Bigr]^{a_1 \cdots a_n}_{\mu_1 \cdots \mu_n}
&= \frac{v_{\mu_n}}{k_n \cdot v + i \epsilon} \,
   \frac{v_{\mu_{n-1}}}{(k_n + k_{n-1}) \cdot v + i \epsilon}
   \cdots
   \frac{v_{\mu_1}}{\rev{(k_n + \cdots + k_1)} \cdot v + i \epsilon}
\nn \\[0.1em]
& \quad \times
    \bigl( t^{a_1} \cdots t^{a_n} \bigr)_{j i} \,,
\end{align}
where $W^{\dagger (n)}$ is defined from the perturbative expansion of $W^\dagger$ as in \eqref{WL-expand}.  Note that $W^{\dagger (n)}(k)$ is the Fourier transform \eqref{fourier-def} of $W^{\dagger (n)}(x)$, i.e.\ the Fourier exponential is still $e^{i k \cdot x}$.  For simplicity, we will refer to  $\bar{\mathcal{E}}$ as ``conjugate eikonal line''.  A graphical representation with $n = 2$ is given in figure~\ref{fig:eikonal_order_2}.  \rev{Note that the sign of $i \epsilon$ is the same in the denominators of \eqref{eqn:eikonal_order_n} and \eqref{eqn:adj_eikonal_order_n}.  It is a consequence of the lower limit $-\infty$ for the $s$ integration in the Wilson line \eqref{eqn:wilson_line_def} and in its complex conjugate.}

%%%%%%%%%%%%%%%%%%

\subsection{Decoupling of collinear gluons}
\label{sec:collinear}

The decoupling of collinear gluons from the hard scattering subprocesses leads from the graphs in figure~\ref{fig:leading_regions} to those in figure~\ref{fig:cross_section_collinear}.  In terms of operators, this corresponds to replacing the quark and antiquark fields in the definition of the initial collinear factors according to
\begin{align}
   q_i (x)       &\to W_{\! v_A,\, i j} (x) \, q_j (x),
\nn \\
   \bar{q}_i (x) &\to \bar{q}_j (x) \, W_{\! v_A,\, j i}^\dagger (x)
\end{align}
for right-collinear partons, where $v_A = (v_A^+, v_A^-, \tvec{0})$ is a spacelike auxiliary vector with either $v_A^- \sim - v_A^+ > 0$ or $v_A^- \gg - v_A^+ > 0$.
If only gluons are exchanged between a hard and a collinear subgraph, then exactly one of them carries transverse polarisation, whilst all others are longitudinal and can be decoupled (contributions with more than one transverse gluons are power suppressed).  A transverse gluon in a right-collinear factor is represented by a gluon field strength operator $G_{\mu\nu}^a$.  After decoupling of the longitudinal gluons, this operator is dressed by an adjoint Wilson line as
\begin{align}
   G_{\mu\nu}^{a} (x) &\to W_{\! v_A}^{a b} (x) \, G_{\mu\nu}^{b} (x) \,.
\end{align}
A corresponding discussion holds for left-collinear factors, with an auxiliary vector $v_B$ and interchanged roles of plus and minus components.  For definiteness, we restrict our discussion to fundamental Wilson lines in the following.  The extension to adjoint Wilson lines is straightforward.

We will discuss the decoupling of soft gluons at the level of the process amplitude rather than the cross section.  A right-collinear subgraph with a quark-antiquark pair and $n$ soft gluons in the amplitude is then given by\footnote{We make no difference between colour indices as sub- or superscripts.}
\begin{align}
\label{JAX-def}
& \Bigl[ J_{A X}^{i_1 i_2} \bigl( l_1, l_2; \{ \ell_n \} \bigr)
  \Bigr]^{a_1 \dots a_n}_{\mu_1 \dots \mu_n}
= \rev{\Biggl[ \ms \prod_{k=1}^{n} \, (i \ell_k^2) \Biggr]} \,
   \Bra{ X \rule{0pt}{1.05em} }
   T A_{\mu_1}^{a_1} (- \ell_1) \dots A_{\mu_n}^{a_n} (- \ell_n) \;
\nn \\[0.2em]
& \qquad \times
   \Gamma_1 \ms \bigl[ W_{\! A}^{}\, q \bigr]_{i_1}\!(l_1) \;
      \bigl[ \bar{q} \ms W^\dagger_{\!\bs A} \ms \bigr]_{i_2}\!(l_2) \,
      \Gamma_2 \Ket{ \rule{0pt}{1.05em}P }
  - \{ \text{subtraction terms} \} \,,
\end{align}
where the ``subtraction terms'' remove the contribution from regions of soft momenta, as was mentioned in point \ref{approx} of section~\ref{sec:overview} and will be expanded on in the next section.
$P$ is the incoming proton and $X$ the set of spectator partons going into the final state, and $T$ denotes time ordering, as appropriate for operators in an amplitude.  The specific form of the Dirac matrices $\Gamma_{1,2}$ is not relevant in our context.  For brevity we write $W_{\! A}$ instead of $W_{v_A}$ for a Wilson line along $v_A$ and define $\bigl[\ms W_{\! A} \, q \ms\bigr](l)$ as the Fourier transform \eqref{fourier-def} of $W_{\! A}(x)\, q(x)$, which is of course equal to the momentum-space convolution of $W_{\! A}(k)$ and $q(l - k)$.  The same holds for $[\ms \bar{q} \, W_{\!\bs A}^\dagger \ms](l)$.  Notice that in $J_{A X}$ the collinear momenta $l_1$ and $l_2$ are taken as outgoing, whilst the soft momenta $\ell_{i}$ are incoming.
The product of $(i \ell_k^2)$ in \eqref{JAX-def} removes the soft gluon propagators, which are contained in the soft subgraph.  The latter is defined by
\begin{align}
\label{SX-def}
& \Bigl[ S_{Z}\bigl( \{ \ell_n \}, \{ \bar{\ell}_{\bar{n}} \} \bigr)
  \Bigr]_{\mu_1 \ldots \mu_n, \, \nu_1 \ldots \nu_{\bar{n}}}
        ^{a_1 \ldots a_n, \, b_1 \ldots b_{\bar{n}}}
\nn \\
& \qquad
=  \Braket{ Z | T A_{\mu_1}^{a_1} (\ell_1) \dots A_{\mu_n}^{a_n} (\ell_n) \,
   A_{\nu_1}^{b_1} (\bar{\ell}_1) \dots
     A_{\nu_{\bar{n}}}^{b_{\bar{n}}} (\bar{\ell}_{\bar{n}})| 0 }
\end{align}
for the process amplitude, \rev{where $Z$ denotes} the set of partons going across the final state cut.  We note that both $J_{A X}$ and $S_{Z}$ are defined such that they include an overall delta function for momentum conservation.

The combination of soft and collinear factors to the left of the final state cut in figure~\ref{fig:cross_section_collinear} is given by
\begin{align}
\label{lum-X-def}
& \mathcal{L}_{X Y Z}^{i_1 i_2 \, \bar{\imath}_1 \bar{\imath}_2 \,
  (n, \bar{n})} (q_1, q_2)
= \frac{1}{n! \, \bar{n}!} \,
  \int \! \fourmomdiff{l_1} \, \fourmomdiff{\bar{l}_1} \, \fourmomdiff{l_2} \, \fourmomdiff{\bar{l}_2} \;
  \prod_{k = 0}^n \, \prod_{\bar{k} = 0}^{\bar{n}}
  \int \! \fourmomdiff{\ell_k} \fourmomdiff{\bar{\ell}_{\bar{k}}} \,
\nn \\
&\quad \times
  \twopif \, \deltaf \left( q_1 - l_1 - \bar{l}_1 \right) \, \twopif \, \deltaf \left( q_2 - l_2 - \bar{l}_2 \right)
  \Bigl[ J_{A X}^{i_1 i_2} \bigl( l_1, l_2; \{ \tilde{\ell}_n \} \bigr)
  \Bigr]^{a_1 \dots a_n}_{\mu_1 \dots \mu_n} \;
\nn \\[0.2em]
&\quad \times
  \Bigl[ S_{Z}^{} \bigl( \{ \ell_n \}, \{ \bar{\ell}_{\bar{n}} \} \bigr)
  \Bigr]^{\mu_1 \ldots \mu_n, \, \nu_1 \ldots \nu_{\bar{n}}}
        _{a_1 \ldots a_n, \, b_1 \ldots b_{\bar{n}}} \;
  \Bigl[ J_{B Y}^{\bar{\imath}_1 \bar{\imath}_2}
  \bigl( \bar{l}_1, \bar{l}_2; \{ \tilde{\bar{\ell}}_{\bar{n}} \} \bigr)
  \Bigr]^{b_1 \dots b_{\bar{n}}}_{\nu_1 \dots \nu_{\bar{n}}} \,,
\end{align}
where $q_1$, $q_2$ are the momenta of the produced gauge bosons.  The left-collinear factor $J_{B Y}$ is defined in analogy to \eqref{JAX-def}.  Soft gluon momenta entering collinear factors are approximated according to \eqref{eqn:soft_approximation_momenta}.  In the term with $n = \bar{n} = 0$ we have a collinear factor $J_{A X}^{i_1 i_2}(l_1, l_2)$, defined as in \eqref{JAX-def} without any external gluon fields, its analogue  $J_{B Y}^{j_1 j_2}(\bar{l}_1, \bar{l}_2)$, and a trivial factor $S_{Z} = 1$.

It is understood that all gluon momenta $\ell_{k}$ and $\bar{\ell}_{\bar{k}}$ in \eqref{lum-X-def} are central soft, because a deformation out of the Glauber region has already been performed as discussed in section~\ref{sec:overview}.  One can therefore insert Grammer-Yennie factors as specified in \eqref{eqn:grammer_yennie_def}:
\begin{align}
\label{ASB-GY}
& \Bigl[ J_{A X}^{i_1 i_2} \bigl( l_1, l_2; \{ \tilde{\ell}_n \} \bigr)
  \Bigr]^{a_1 \dots a_n}_{\mu_1 \dots \mu_n} \;
  \Bigl[ S_{Z}^{} \bigl( \{ \ell_n \}, \{ \bar{\ell}_{\bar{n}} \} \bigr)
  \Bigr]^{\mu_1 \ldots \mu_n, \, \nu_1 \ldots \nu_{\bar{n}}}
        _{a_1 \ldots a_n, \, b_1 \ldots b_{\bar{n}}} \;
  \Bigl[ J_{B Y}^{\bar{\imath}_1 \bar{\imath}_2}
  \bigl( \bar{l}_1, \bar{l}_2; \{ \tilde{\bar{\ell}}_{\bar{n}} \} \bigr)
  \Bigr]^{b_1 \dots b_{\bar{n}}}_{\nu_1 \dots \nu_{\bar{n}}}
\nn \\[0.6em]
& \quad
\approx
  Y_{R}^{\mu_1 \alpha_1} (\tilde{\ell}_1) \, \dots \,
  Y_{R}^{\mu_n \alpha_n} (\tilde{\ell}_n) \,
  \Bigl[ J_{A X}^{i_1 i_2} \bigl( l_1, l_2; \{ \tilde{\ell}_n \} \bigr)
  \Bigr]^{a_1 \dots a_n}_{\alpha_1 \dots \alpha_n} \;
  \Bigl[ S_{Z}^{} \bigl( \{ \ell_n \}, \{ \bar{\ell}_{\bar{n}} \} \bigr)
  \Bigr]_{\mu_1 \ldots \mu_n, \, \nu_1 \ldots \nu_{\bar{n}}}
        ^{a_1 \ldots a_n, \, b_1 \ldots b_{\bar{n}}} \;
\nn \\[0.5em]
& \qquad \times
  Y_{L}^{\nu_1 \beta_1} (\tilde{\bar{\ell}}_1) \, \dots \,
  Y_{R}^{\nu_n \beta_n} (\tilde{\bar{\ell}}_{\bar{n}}) \,
  \Bigl[ J_{B Y}^{\bar{\imath}_1 \bar{\imath}_2}
  \bigl( \bar{l}_1, \bar{l}_2; \{ \tilde{\bar{\ell}}_{\bar{n}} \} \bigr)
  \Bigr]^{b_1 \dots b_{\bar{n}}}_{\beta_1 \dots \beta_{\bar{n}}} \,.
\end{align}

A remark on combinatorics is in order here.  The Green functions defined by operator matrix elements in \eqref{JAX-def} and \eqref{SX-def} both contain a sum over all combinations of coupling the gluons with momenta $\ell_1, \ldots \ell_n$ to the internal lines of the Green function.
Contracting them (and integrating over the gluon momenta) one over-counts by a factor of $n!$ the combinations of connecting all internal lines of $J_{A X}$ with all internal lines of $S_{X}$ by $n$ gluons.  This factor $n!$ is therefore divided out in \eqref{lum-X-def}.  The same argument holds of course for the factor $\bar{n}!$.

The amplitude for the double Drell-Yan process is obtained by multiplying \eqref{lum-X-def} with the hard subgraphs.  One then needs to sum over all $n$ and $\bar{n}$,  over all relevant combinations of quarks and antiquarks that can be exchanged between the collinear and hard subgraphs, and over their colour and polarisation.  In the next section, we will show how to decouple the soft gluons exchanged between subgraphs $J_A$ and $S$ in the amplitude.  Applying the same argument to the conjugate amplitude and to the soft gluons exchanged between $J_B$ and $S$, one arrives at the expression represented in figure~\ref{fig:cross_section_factorised}.  Further steps towards the final factorisation formula will be discussed in section~\ref{sec:final_result}.

\section{Soft gluon decoupling}
\label{sec:proof}

We now show how to decouple soft gluons from collinear factors in double parton scattering.  The steps in this section closely follow the arguments given for single parton scattering in~\cite{Collins:1988ig}.  The main differences between our treatment and the one in~\cite{Collins:1988ig}
are either due to the difference in kinematic approximations mentioned below \eqref{eqn:soft_approximation_momenta} or due to the fact that for DPS the collinear subgraphs have two external parton lines instead of one, which in particular requires one to keep track of colour indices.  Another difference between the two treatments is pointed out below.

On the r.h.s.\ of \eqref{ASB-GY}, the gluon polarisation indices of $J_{A X}$ are all contracted with the corresponding approximated gluon momenta, i.e.\ we have a factor of the form
\begin{align}
\label{JAX-contracted}
\ell_1^{\mu_1} \dots \ell_{n}^{\mu_n} \;
\Bigl[ J_{A X}^{i_1 i_2} \bigl( l_1, l_2; \{ \ell_n \} \bigr)
  \Bigr]^{a_1 \dots a_n}_{\mu_1 \dots \mu_n} \,.
\end{align}
For better legibility, we omit the tilde for the approximation \eqref{eqn:soft_approximation_momenta} of soft momenta in the remainder of this section, i.e.\ it is understood that
\begin{align}
\label{soft-approx-short}
\ell_k^+ &= 0 & & \text{~for~} 1 \le k \le n \,.
\end{align}
The form \eqref{JAX-contracted} allows us to use the Ward identity
\begin{align}
\label{full-ward}
   \ell_1^{\mu_1} \dots \ell_{n}^{\mu_n} \;
   \Bra{ X \rule{0pt}{1.05em} }
   T A_{\mu_1}^{a_1} (- \ell_1) \dots A_{\mu_n}^{a_n} (- \ell_n) \;
   \Gamma_1 \bigl[ W_{\! A}^{}\, q \bigr]_{i_1}(l_1) \;
      \bigl[ \bar{q} \ms W^\dagger_{A} \bigr]_{i_2}(l_2) \Gamma_2
   \Ket{ \rule{0pt}{1.05em}P }
 & = 0 \,.
\end{align}
This requires, however, a careful discussion because $J_{A X}$ is a \emph{collinear} subgraph and not a full Green function as appears in \eqref{full-ward}.  This means that an incoming soft gluon in \eqref{JAX-contracted} couples \emph{directly} to an \emph{internal} line inside $J_{A X}$, which by construction carries a collinear momentum.  This has two implications.
\begin{enumerate}
\item \label{case:fixed-ward} We need the Ward identity not for the full Green function (where internal loop momenta are integrated over the full phase space) but for the case in which the internal momenta in the Green function are all fixed.  To be more precise, consider a particular collinear graph with external lines collinear and on shell, and with all internal momenta collinear and fixed.  Connecting $n$ external gluons to this graph in all possible ways and contracting  gluon polarisation indices with the corresponding gluon momenta gives zero.  This form of Ward identities can be proven in a diagrammatic approach
\cite{tHooft:1971akt,tHooft:1972qbu}.

An additional complication comes from the presence of the operators $W_{\!\bs A}^{}\, q$ and $\bar{q} \ms W^\dagger_{\!\bs A}$ in \eqref{full-ward}, which are constructed from nonlocal products of fields in position space and do not correspond to on-shell external collinear lines.  This case was not considered in the work just quoted.  In the appendix we prove  \eqref{full-ward} for the full Green function using functional methods.  An extension to the case of graphs with fixed internal momenta is beyond the scope of this work.
\item The Green function in \eqref{full-ward} contains graphs not included in \eqref{JAX-contracted}, namely graphs
\begin{enumerate}
\item \label{case:soft-self} in which two or more soft gluons interact with each other before attaching to an internal collinear line,
\item \label{case:WL} or in which one or more soft gluon couple to the eikonal lines generated by the operators $W_{\!\bs A}^{}$ or $W^\dagger_{\!\bs A}$, either directly or after interacting with each other,
\end{enumerate}
or in which the two previous cases are combined.  Applying a Ward identity to  \eqref{JAX-contracted} thus schematically gives
\begin{align}
\label{JAX-ward}
& \ell_1^{\mu_1} \dots \ell_{n}^{\mu_n} \;
\Bigl[ J_{A X} \bigl( l_1, l_2; \{ \ell_n \} \bigr)
  \Bigr]_{\mu_1 \dots \mu_n}
\nn \\[0.2em]
&\quad
  = - \ell_1^{\mu_1} \dots \ell_{n}^{\mu_n} \;
   \Bigl[\, \text{graphs (a)} + \text{graphs (b)} + \text{combinations of (a) and (b)} \,\Bigr] \,.
\end{align}
\end{enumerate}

\begin{figure}
\centering
\subfigure[\label{fig:gluon-self-good}]{\includegraphics[height=0.17\textwidth]{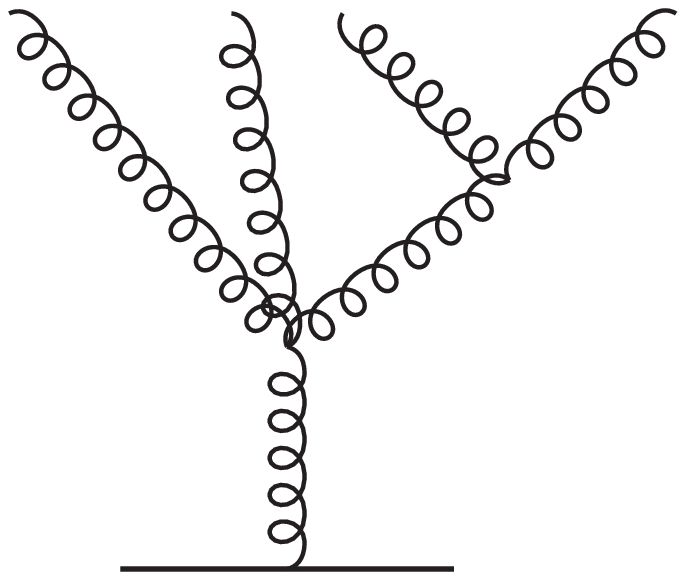}}
\hspace{5em}
\subfigure[\label{fig:gluon-self-bad}]{\includegraphics[height=0.17\textwidth]{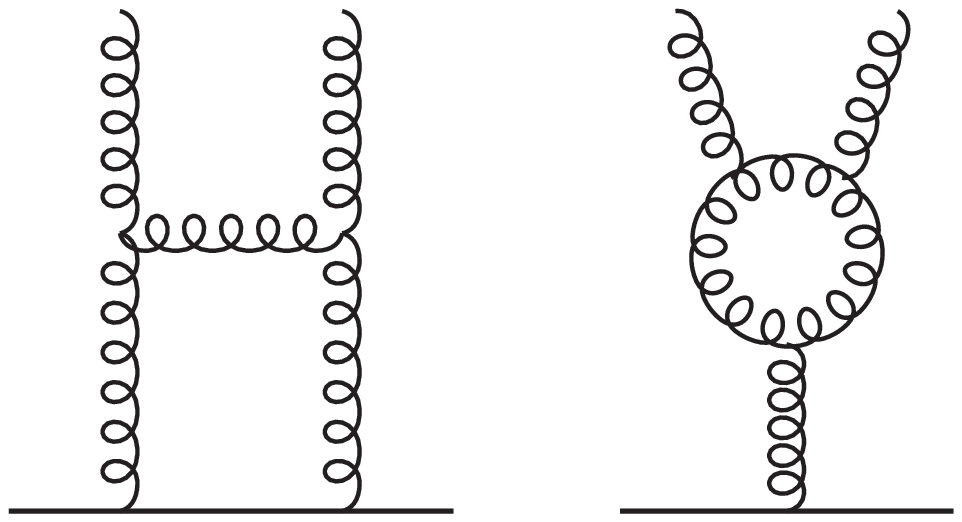}}
\caption{Example graphs for $n \to m$ gluon transitions of order $g^{n-m}$ (a) and of higher order (b).}
\end{figure}

The treatment of soft gluon interactions is simplified by the following considerations.  Take the collinear graph without external soft gluons mentioned in point \ref{case:fixed-ward} to be of order $g^c$ in the strong coupling, and restrict the corresponding Ward identity for $n$ external gluon attachments to the smallest possible order, which is $g^{c + n}$.  Interactions between the $n$ incoming gluons can result in $m$ soft gluons, each of which then couples to a collinear line.  The latter gives a factor $g^m$, so that the interactions among the soft gluons are of order $g^{n-m}$.  It immediately follows that $m \le n$.  One finds that transitions from $n$ to $m$ gluons of order $g^{n-m}$ are only realised by tree graphs with sequential transitions from $2 \to 1$ or from $3 \to 1$ gluons, as shown in figure~\ref{fig:gluon-self-good}.  Any further interactions, as in figure~\ref{fig:gluon-self-bad} come along with further powers of $g$ and are hence excluded.  The result is that the graphs with soft gluon interactions are restricted to a disconnected set of transitions from $n_i \to 1$ gluons, with $n_i \ge 2$ and $\sum_i n_i \le n$.  We note that the proof in \cite{Collins:1988ig} did not restrict the order of the coupling and used a different argument to eliminate soft gluon interactions involving loops.

%%%%%%%%%%%%%%%%%%%%%%%%%%%%%%%

\subsection{Leading power approximations}
\label{sec:approximations}

The r.h.s.\ of \eqref{JAX-ward} admits two types of approximation, valid at leading order in $\Lambda /Q$, which we now discuss in turn.

\subsubsection*{Soft gluon interactions}

Consider a graph with gluon self interactions, as characterised in the previous subsection.  We denote the tree-level graphs for $m \to 1$ gluon transitions by $V^{(m)}( \{ \ell_m \} )$, where the argument specifies the $m$ incoming gluon momenta.  The outgoing momentum is of course $\ell = \sum_{k=1}^{m} \ell_k$.  The propagators of the incoming gluons are truncated in $V^{(m)}$, but not the one of the outgoing gluon.

\begin{figure}
\centering
\includegraphics[scale=0.6]{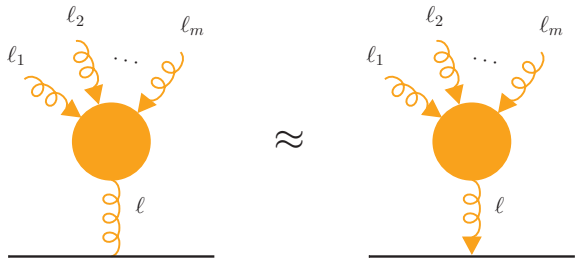}
\caption{\label{fig:GY-insert} Graphical representation of the approximation~\protect\eqref{GY-insert}.  The line at the bottom can be either a right-collinear parton or an eikonal line along $v_A$.  Here and in the following, an arrow on a gluon line represents the insertion of a Grammer-Yennie factor \protect\eqref{eqn:grammer_yennie_def}, with the arrow on the side of the momentum vector.  Round blobs denote the tree-level vertices $V^{(m)}$ specified in the text.}
\end{figure}

In the graphs on the r.h.s.\ of \eqref{JAX-ward}, a $m\to 1$ vertex is contracted with the incoming gluon momenta.  Due to Lorentz invariance, this contraction can be written as
\begin{align}
\label{soft-decomp}
\Bigl[ \tilde{V}^{(m)} \bigl( \{ \ell_m \} \bigr) \Bigr]_{\nu}
 = \ell_1^{\mu_1} \dots \ell_{m}^{\mu_{m}} \,
   \Bigl[ V^{(m)} \bigl( \{ \ell_m \} \bigr)
     \Bigr]_{\mu_1 \dots \mu_{m} \ms \nu}
 &= \sum_{k=1}^{m} (\ell_{k})_{\nu}^{} \;
    F^{(m,k)} \bigl( \{ \ell_m \} \bigr) \,,
\end{align}
where $F^{(m,k)}$ is a set of scalar functions and we drop colour indices for brevity.  The index $\nu$ refers to the outgoing gluon and is to be contracted with the Lorentz index of a collinear subgraph (for graphs a) or of an eikonal line along $v_A$ (for graphs b).  In both cases, one only needs the minus component of $\tilde{V}^\nu$ in the contraction.  In the latter case this is because $\ell_k^{} \cdot v_{A}^{} =  \ell_k^-\, v_A^+$ for momenta satisfying \eqref{soft-approx-short}, and in the former case this follows from the same argument as in the original Grammer-Yennie approximation \eqref{soft-GY}.  Using the form \eqref{soft-decomp}, one readily obtains
\begin{align}
\label{GY-insert}
\Bigl[ \tilde{V}^{(m)} \bigl( \{ \ell_m \} \bigr) \Bigr]^-
&= \sum_{k=1}^{m} \ell_k^{-} \, F^{(m,k)} \bigl( \{ \ell_m \} \bigr) \,
  \frac{v_R^+ \, \ell^{-}}{v_R \cdot \ell + i \epsilon}
 = \Bigl[ \tilde{V}^{(m)} \bigl( \{ \ell_m \} \bigr)
     \Bigr]_{\mu} \,
  Y_R^{\mu -} (\ell) \,.
\end{align}
This means that, because the incoming gluons in a $m \to 1$ gluon vertex come with a Grammer-Yennie factor, one may insert the same factor again for the final gluon.  We represent this result graphically in figure~\ref{fig:GY-insert}.

We note that the previous argument drastically simplifies if one can use the soft momentum approximation of Collins et al.\ \cite{Collins:1988ig}, described after \eqref{eqn:soft_approximation_momenta}.  The invariant functions $F^{(m,k)}$ in \eqref{soft-decomp} are homogeneous polynomials in the scalar products of soft momenta, which readily follows from the Feynman rules for the three- and four-gluon vertices.  In the approximation of \cite{Collins:1988ig} one keeps only the minus components of these momenta, so that the scalar products and hence all $F^{(m,k)}$ are zero.  Soft gluon interactions need then not be considered at all.  However, as already noted, this approximation can be used for SPS but not for DPS graphs.

%%%%%%%%%%%%%%%%%%%%%%%%%%%%%%%%%%%%%%%%

\subsubsection*{Soft gluon attachments to eikonal lines}

\begin{figure}
\centering
\subfigure[\label{fig:soft-to-WL-good}]{\includegraphics[scale=0.56]{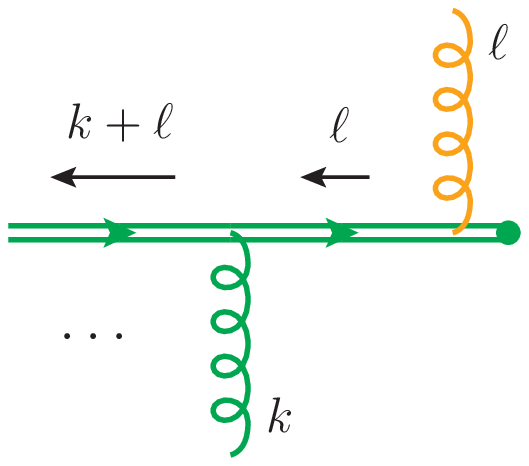}}
\hspace{5em}
\subfigure[\label{fig:soft-to-WL-bad}]{\includegraphics[scale=0.56]{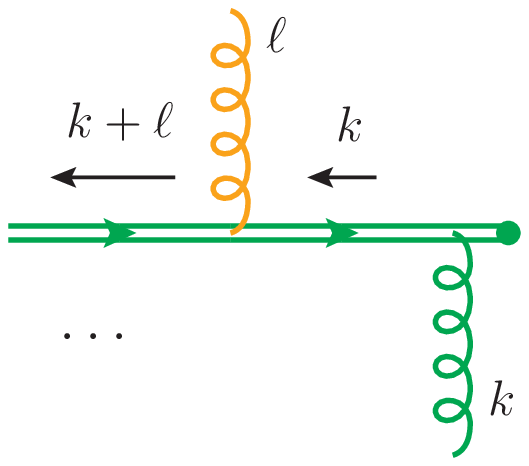}}
\caption{The two possibilities of attaching a soft and a collinear gluon to an eikonal line.}
   \label{fig:soft_glu_to_wilson}
\end{figure}

We now turn to soft gluon attachments to eikonal lines generated by the operators $W_{\! A}^{}$ or $W^\dagger_{A}$ in \eqref{JAX-def}.  It is easy to see that only those graphs are dominant in which first all soft gluons attach to the eikonal line and then all collinear ones, as illustrated in figure~\ref{fig:soft-to-WL-good}.  To see this, we note that for collinear momenta $k_i$ one has $v_A^{} \cdot k \approx v_A^- \, k^+$ and for soft momenta $\ell$ approximated as in \eqref{soft-approx-short} one has $v_A^{} \cdot \ell^{} = v_A^+ \, \ell^-$, so that
\begin{align}
\label{eik-compare}
| v_A \cdot k \ms | & \gg | v_A \cdot \ell | \,.
\end{align}
Comparing the eikonal propagators for the two orderings of attaching a soft and a collinear gluon in figure~\ref{fig:soft-to-WL-good} and \ref{fig:soft-to-WL-bad}, we thus obtain
\begin{align}
\biggl| \frac{1}{v_A \cdot \ell} \; \frac{1}{v_A \cdot (k + \ell)}
  \biggr| \gg \biggl|
  \frac{1}{v_A \cdot k} \; \frac{1}{v_A \cdot (k + \ell)} \biggr| \,.
\end{align}
The argument readily generalises to several collinear and soft momenta.

Using \eqref{eik-compare}, one can further split the dominant terms into two eikonal factors.  For one soft and one collinear gluon, one has
\begin{align}
\Bigl[ \mathcal{E}^{(2)}_{A, i j} (k, \ell)
   \Bigr]_{a b}^{\mu \nu}
&= \frac{- v_{A}^{\nu}}{v_A \cdot \ell + i\epsilon} \, t^b_{i h} \;
   \frac{- v_{A}^{\mu}}{v_A \cdot (\ell + k) + i\epsilon} \, t^{a}_{h j}
\approx  \frac{- v_{A}^{\nu}}{v_A \cdot \ell + i\epsilon} \; t^b_{i h} \;
   \frac{- v_{A}^{\mu_n}}{v_A \cdot k + i\epsilon} \, t^{a}_{h j}
\nn \\[0.2em]
&= \Bigl[ \mathcal{E}^{(1)}_{A, i h} (\ell) \Bigr]_{b}^{\nu} \;
   \Bigl[ \mathcal{E}^{(1)}_{A, h j} (k) \Bigr]_{a}^{\mu} \,.
\end{align}
In analogy, one obtains
\begin{align}
\label{eqn:soft_approximation_wilson_lines}
\Bigl[ \mathcal{E}^{(m+n)}_{A, i j} \bigl(\{ k_m \}, \{ \ell_n \} \bigr)
   \Bigr]_{a_1 \dots a_m b_1 \dots b_n}^{\mu_1 \dots \mu_m \nu_1 \dots \nu_n}
&\approx
   \Bigl[ \mathcal{E}^{(n)}_{A, i h} \bigl( \{ \ell_n \} )
   \Bigr]_{b_1 \dots b_n}^{\nu_1 \dots \nu_n} \;
   \Bigl[ \mathcal{E}^{(m)}_{A, h j} \bigl(\{ k_m \} \bigr)
   \Bigr]_{a_1 \dots a_m}^{\mu_1 \cdots \mu_m} \,,
\end{align}
as well as the relation
\begin{align}
\label{eqn:soft_approximation_adj_WL}
   \Bigl[ \bar{\mathcal{E}}^{(m+n)}_{A, j i} \bigl(\{ k_m \},
     \{ \ell_n \}\bigr)
   \Bigr]_{a_1 \dots a_m b_1 \dots b_n}^{\mu_1 \dots \mu_m \nu_1 \dots \nu_n}
&\approx
   \Bigl[ \bar{\mathcal{E}}^{(m)}_{A, j h} \bigl(\{ k_m \} \bigr)
   \Bigr]_{a_1 \dots a_m}^{\mu_1 \cdots \mu_m} \;
   \Bigl[ \bar{\mathcal{E}}^{(n)}_{A, h i} \bigl( \{ \ell_n \}\bigr)
   \Bigr]_{b_1 \dots b_n}^{\nu_1 \dots \nu_n}
\end{align}
for conjugate eikonal lines.  In both cases one can replace the original expression with the product of two eikonal lines, one with only soft momenta and one with only collinear ones.

%%%%%%%%%%%%%%%%%%%%%%%%%%%%%%%%%%%%%%%%%%%%%%%%%

\subsection{Decoupling of one soft gluon}

Let us now consider the case where a single soft gluon couples to the collinear subgraph.  This is the base case for the inductive proof in section~\ref{sec:multiple_gluon_attachments}.  We first treat single parton scattering, and then show how it generalises to DPS.  It will become clear that generalisation to $N$ hard scatters is straightforward.

%%%%%%%%%%%%%%%%%%%%%%%%%%%%%%%

\subsubsection{Single parton scattering}

\begin{figure}
\centering
\includegraphics[scale=0.53, trim=0 0 480 0, clip]{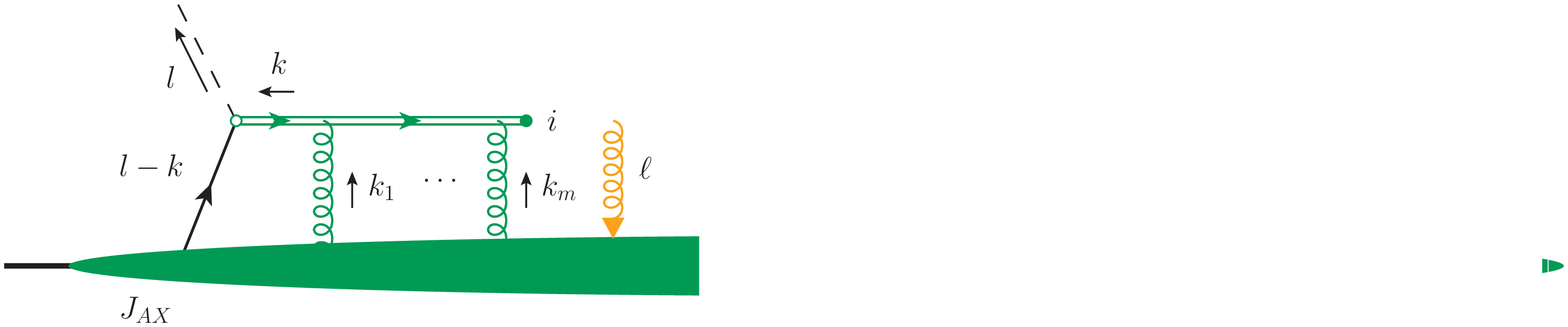} \\[1em]
\includegraphics[scale=0.53, trim=0 0 480 0, clip]{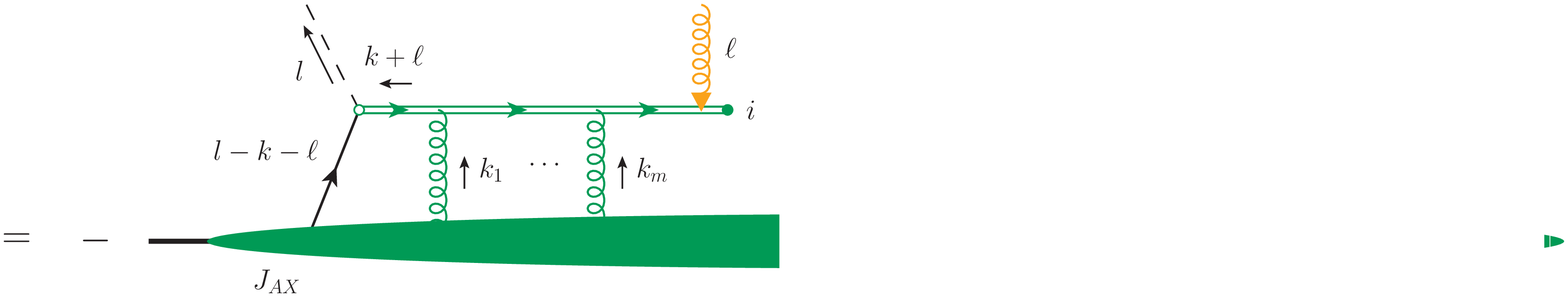} \\[0.5em]
\includegraphics[scale=0.53, trim=0 0 480 0, clip]{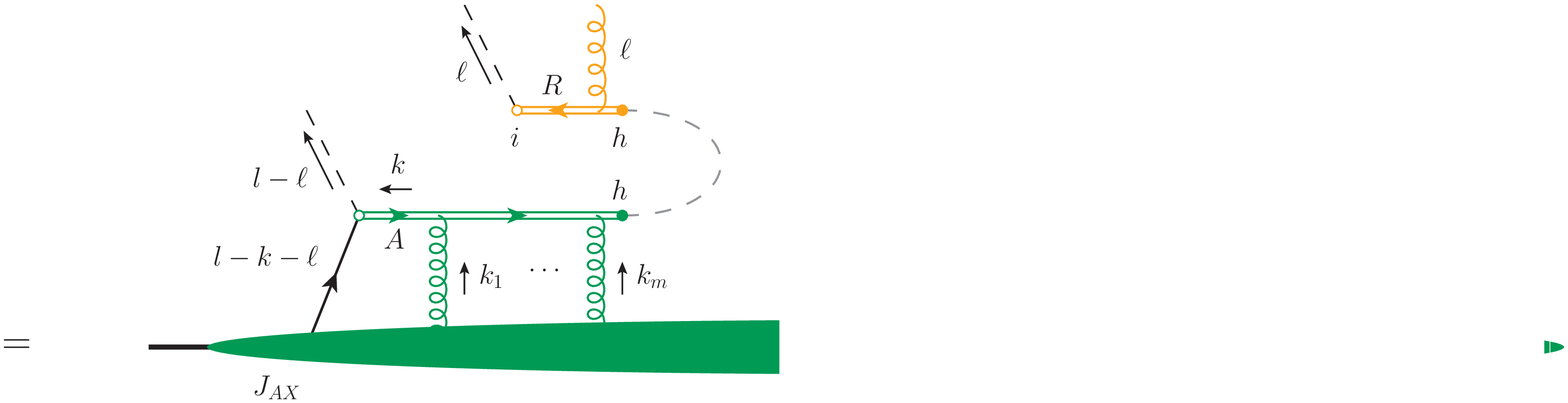}
\caption{\label{fig:sps_one_gluon} Graphical representation for the decoupling of a single soft gluon from a collinear graph in the amplitude of an SPS process.
In the last row, eikonal lines point along $v_A$ and $v_R$ as indicated.  The dashed line between the two eikonal lines is drawn to visualise the contraction of colour indices.}
\end{figure}

The collinear factor for SPS in the amplitude is defined as in \eqref{JAX-def}, but with one parton operator instead of two, which for definiteness we take as $W_{\! A} \, q$.  Starting point of our consideration is the expression
\begin{equation} \label{eqn:one_gluon_sps}
   Y_{R}^{\mu\nu} (\ell) \, \Bigl[ J_{A X}^{i} (l; \ell) \Bigr]^{b}_{\nu} \,,
\end{equation}
which includes the Grammer-Yennie factor~\eqref{eqn:grammer_yennie_def} and is illustrated in the first line of figure~\ref{fig:sps_one_gluon}.  Since $Y_{R}^{\mu\nu} (\ell) \propto \ell^\nu$, we can employ the SPS analogue of Ward identity \eqref{JAX-ward} for $n=1$.  The only graph contributing to the r.h.s.\ of this identity is shown in the second line of the figure and has the soft gluon coupling at the end of the eikonal line along $v_A$.

Let us consider a graph with $m$ collinear gluons coupling to the eikonal line along $v_A$, in addition to the soft gluon.    Contracting with the Grammer-Yennie factor of the latter, we obtain
\begin{align}
\label{eqn:one_gluon_R}
Y_R^{\mu\nu} (\ell) \,
  \Bigl[ \mathcal{E}^{(m+1)}_{A, i j} \bigl(\{ k_m \}, \ell \bigr)
   \Bigr]^{a_1 \dots a_m b}_{\mu_1 \dots \mu_m \nu}
&\approx \frac{v_R^\mu}{v_R \cdot \ell + i\epsilon} \,
   \frac{- v_A \cdot \ell}{v_A \cdot \ell + i\epsilon} \; t^b_{i h} \,
   \Bigl[ \mathcal{E}^{(m)}_{A, h j} \bigl(\{ k_m \} \bigr)
   \Bigr]^{a_1 \dots a_m}_{\mu_1 \dots \mu_m}
\nn \\[0.4em]
&= {}- \Bigl[ \bar{\mathcal{E}}^{(1)}_{R, i h} (\ell) \bigr) \Bigr]^{\mu}_b \;
   \Bigl[ \mathcal{E}^{(m)}_{A, h j} \bigl(\{ k_m \} \bigr)
   \Bigr]^{a_1 \dots a_m}_{\mu_1 \dots \mu_m} \,.
\end{align}
In the first step, we used the relation \eqref{eqn:soft_approximation_wilson_lines} with $n=1$ and explicitly wrote out $Y_R$ and $\mathcal{E}^{(1)}_A$, and in the second step we used the representation \eqref{eqn:adj_eikonal_order_n} of a conjugate eikonal line.  The reason for taking $- \bar{\mathcal{E}}_R$ rather than $\mathcal{E}_R$ at this point will only become clear when considering more than one soft gluon attachments.  Combining the minus sign in the Ward identity with the one in \eqref{eqn:one_gluon_R}, we obtain
\begin{align}
\label{one_gluon_sps}
Y_{R}^{\mu\nu} (\ell) \, \Bigl[ J_{A X}^{i} (l; \ell)
   \Bigr]^{b}_{\nu}
& \approx
\Bigl[ \bar{\mathcal{E}}^{(1)}_{R, i h} (\ell) \bigr) \Bigr]^{\mu}_b \;
   J_{A X}^{h} (l - \ell)  \,,
\end{align}
where we have summed over all numbers $m = 0,1,\dots$ of collinear gluons in \eqref{eqn:one_gluon_R}.  On the r.h.s.\ of \eqref{one_gluon_sps}, \rev{the eikonal line $\mathcal{E}^{(m)}_{A}$} has become part of  the collinear factor $J_{A X} (l - \ell)$ with no external soft gluon.  The shift of the collinear momentum argument from $l$ to $l - \ell$ follows from momentum conservation, as depicted graphically in figure~\ref{fig:sps_one_gluon}.
The soft gluon is now completely decoupled from the collinear factor, except for the contraction of colour indices shown in the last line of the figure.

%%%%%%%%%%%%%%%%%%%%%%%%%%%%%%%%%%%%%%%%%%%

\subsubsection{Double parton scattering}

\begin{figure}
\centering
\includegraphics[scale=0.455, trim=0 0 20 0, clip]{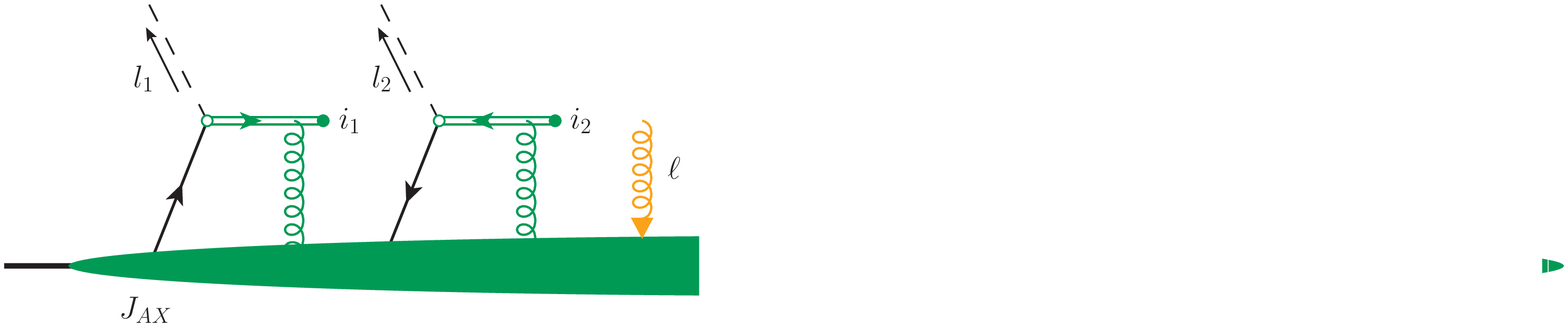} \\[1em]
\includegraphics[scale=0.455]{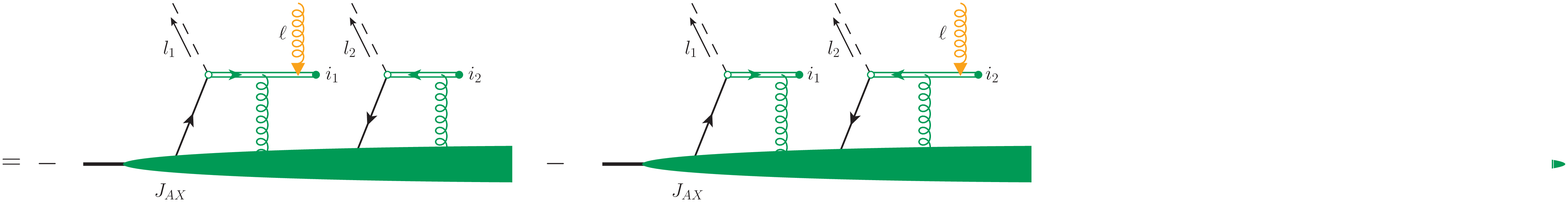} \\[0.5em]
\includegraphics[scale=0.455]{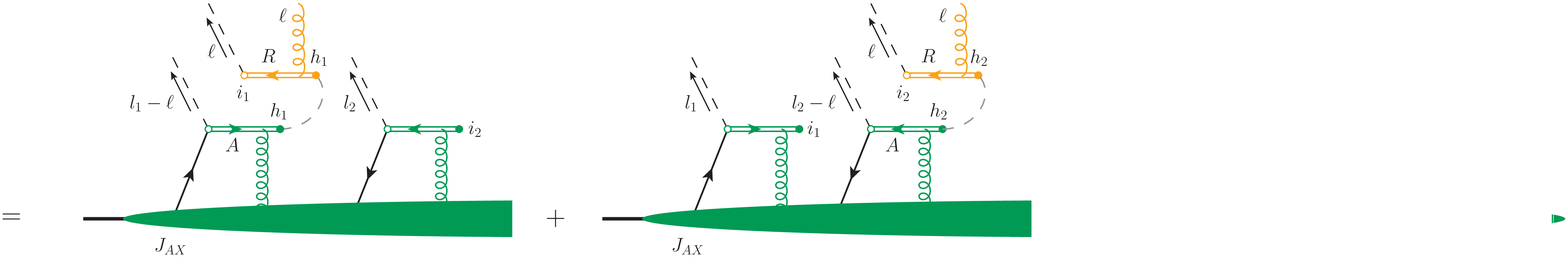}
\caption{Decoupling of a single soft gluon from a collinear graph in the amplitude of a DPS process.}
   \label{fig:one_gluon_attachment_dpd}
\end{figure}

In the case of double parton scattering we have two eikonal lines along $v_A$ in the collinear subgraph, each belonging to one of the partons entering a hard scattering in the overall process.  To be specific, we consider the case of a quark and an antiquark here, so to exhibit the difference between eikonal lines $\mathcal{E}$ and $\bar{\mathcal{E}}$.  A graphical representation of the argument is given in figure~\ref{fig:one_gluon_attachment_dpd}.

Starting point in the DPS case is the expression
\begin{equation}
   Y_{R}^{\mu\nu} (\ell) \, \Bigl[ J_{A X}^{i_1 i_2} (l_1, l_2; \ell)
   \Bigr]^{b}_{\nu} \,,
\end{equation}
and the main difference to SPS is that there are now two graphs of type (a) that appear at the r.h.s.\ of the Ward identity \eqref{JAX-ward}.  If the soft gluon couples to $\mathcal{E}_{\! A}$, we use again \eqref{eqn:one_gluon_R}, and if it couples to $\bar{\mathcal{E}}_{\! A}$, we use its analogue
\begin{align}
\label{eqn:one_gluon_adj_R}
Y_R^{\mu\nu} (\ell) \,
  \Bigl[ \bar{\mathcal{E}}^{(m+1)}_{A, j i} \bigl(\{ k_m \}, \ell \bigr)
   \Bigr]^{a_1 \dots a_m b}_{\mu_1 \dots \mu_m \nu}
&\approx \Bigl[ \bar{\mathcal{E}}^{(m)}_{A, j h} \bigl(\{ k_m \} \bigr)
   \Bigr]^{a_1 \dots a_m}_{\mu_1 \dots \mu_m} \,
   \frac{v_R^\mu}{v_R \cdot \ell + i\epsilon} \,
   \frac{v_A \cdot \ell}{v_A \cdot \ell + i\epsilon} \; t^b_{h i} \,
\nn \\[0.4em]
&= {}- \Bigl[ \bar{\mathcal{E}}^{(m)}_{A, j h} \bigl(\{ k_m \} \bigr)
   \Bigr]^{a_1 \dots a_m}_{\mu_1 \dots \mu_m} \,
   \Bigl[ {\mathcal{E}}^{(1)}_{R, h i} (\ell) \bigr) \Bigr]^{\mu}_b \,,
\end{align}
which is readily obtained from \eqref{eqn:soft_approximation_adj_WL}.  Summing over the two contributing graphs, we obtain
\begin{align}
\label{eqn:one_gluon_factorisation_dps}
Y_{R}^{\mu\nu} (\ell) \, \Bigl[ J_{A X}^{i_1 i_2} (l_1, l_2; \ell)
   \Bigr]^{b}_{\nu}
& \approx
  \Bigl[ \bar{\mathcal{E}}^{(1)}_{R, i_1 h} (\ell) \Bigr]^{\mu}_b \;
  J_{A X}^{h \ms i_2} (l_1 - \ell, l_2)
\nn \\
& \quad
+ J_{A X}^{i_1 h} (l_1, l_2 - \ell) \;
  \Bigl[ {\mathcal{E}}^{(1)}_{R, h \ms i_2} (\ell) \Bigr]^{\mu}_b \,.
\end{align}
On the r.h.s.\ we have collinear factors with no external soft gluons, with different shifts of the collinear momentum arguments arising from momentum conservation.  We note in passing that \eqref{eqn:one_gluon_factorisation_dps} corresponds to equation (3.22) in \cite{Diehl:2011yj}, which was verified for two explicit examples but not generally proven in that work.

%%%%%%%%%%%%%%%%%%%%%%%%%%%%%%%%%%%%%%%%%%%%%%%%%%%%%%%%%%%%%%%%%%%%%%%%

\subsection{Multiple soft gluon attachments}
\label{sec:multiple_gluon_attachments}

The formula for the decoupling of $n$ soft gluons from the collinear graph reads
\begin{align}
 \label{eqn:multi_gluon_factorised}
& Y_{R}^{\mu_1 \nu_1} (\ell_1) \, \dots \, Y_{R}^{\mu_n \nu_n} (\ell_n) \,
\Bigl[ J_{A X}^{i_1 i_2} \bigl( l_1, l_2; \{ \ell_n \} \bigr)
   \Bigr]^{b_1 \dots b_n}_{\nu_1 \dots \nu_n}
= \sum_{ \pi, \sigma_1, \sigma_2 \atop
        |\sigma_1| + |\sigma_2| = n \rule{0pt}{1.2ex} }
  \Bigl[ \bar{\mathcal{E}}^{(|\sigma_1|)}_{R, i_1 h_1}
     \bigl( \pi \{ \ell \}_{\sigma_1} \bigr)
  \Bigr]^{ \pi \{ \mu \}_{\sigma_1}}_{ \pi \{ b \}_{\sigma_1}} \;
\nn \\[0.3em]
& \qquad\quad \times
  J_{A X}^{h_1 h_2} (l_1 - \ell_{\sigma_1}, l_2 - \ell_{\sigma_2}) \;
  \Bigl[ {\mathcal{E}}^{(|\sigma_2|)}_{R, h_2 i_2}
     \bigl( \pi \{ \ell \}_{\sigma_2} \bigr)
  \Bigr]^{ \pi \{ \mu \}_{\sigma_2}}_{ \pi \{ b \}_{\sigma_2}} \,,
\end{align}
where the equality holds up to power corrections.  On the r.h.s.\ the $n$ soft gluon momenta have been partitioned into two sets $\sigma_1$ and $\sigma_2$ with $|\sigma_1|$ and $|\sigma_2|$ elements, respectively.  By $\pi \{ \ell \}_{\sigma_i}$ we denote a particular permutation $\pi$ of the momenta in the set, with $\pi \{ \mu \}_{\sigma_i}$ and $\pi \{ b \}_{\sigma_i}$ being the corresponding permutations of Lorentz and colour indices.  The sum runs over all partitions $\sigma_1$, $\sigma_2$ and over all permutations of their respective members.  The sum of soft momenta in set $i$ is denoted by $\ell_{\sigma_i}$.  For $n=1$, the above expression simplifies to \eqref{eqn:one_gluon_factorisation_dps}, as it must be.

\begin{figure}
\centering
\includegraphics[scale=0.5, trim=0 0 460 0, clip]{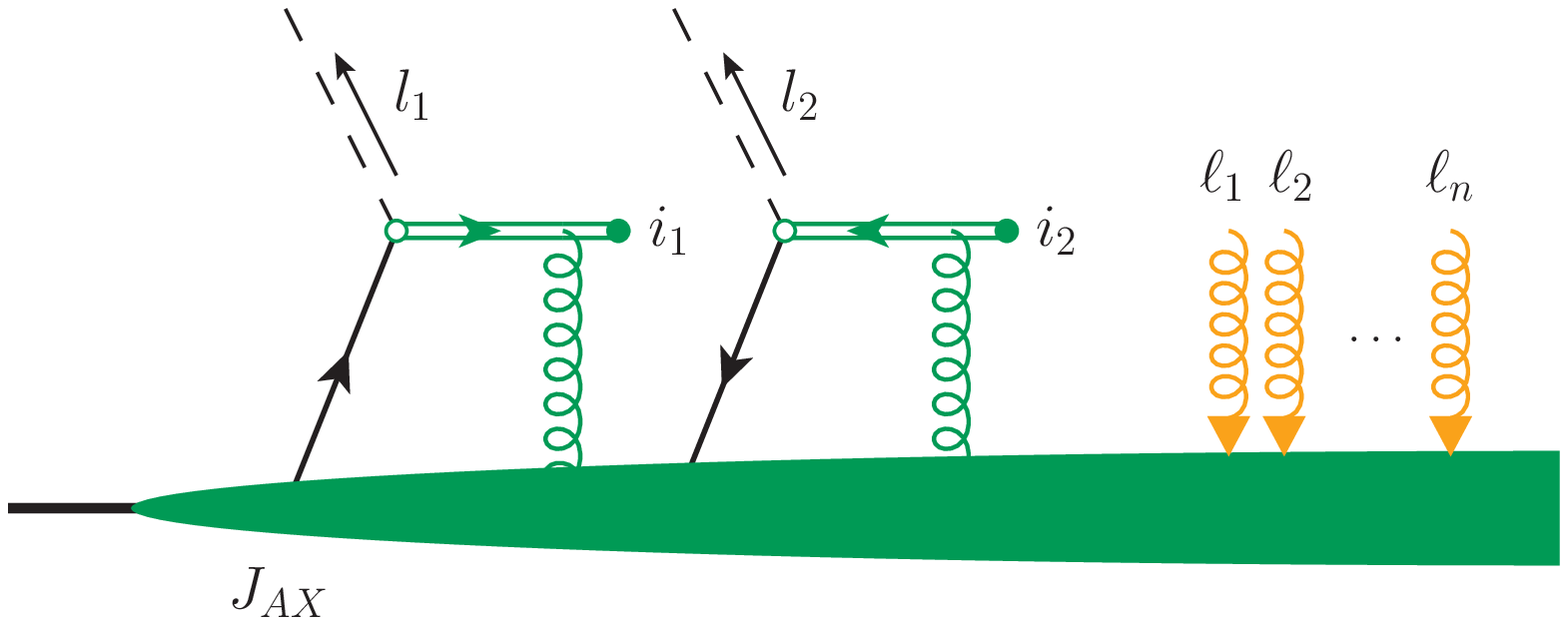} \\[1em]
\includegraphics[scale=0.5, trim=0 0 460 0, clip]{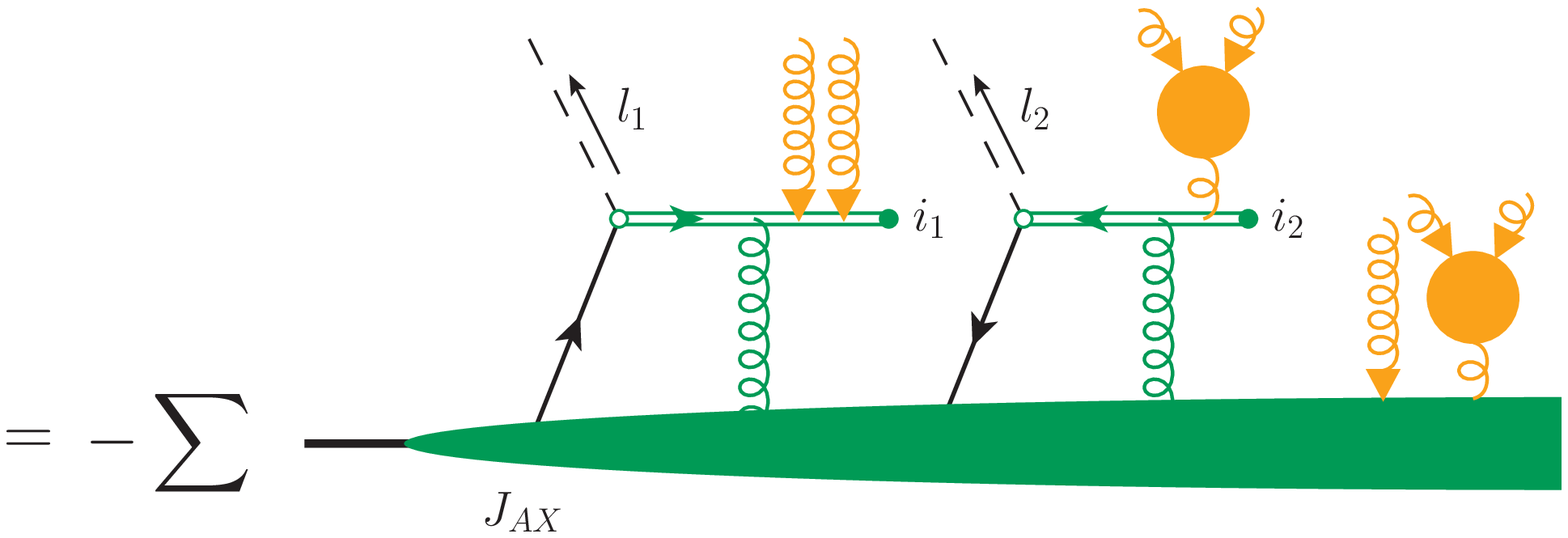} \\[1em]
\includegraphics[scale=0.5, trim=0 0 460 0, clip]{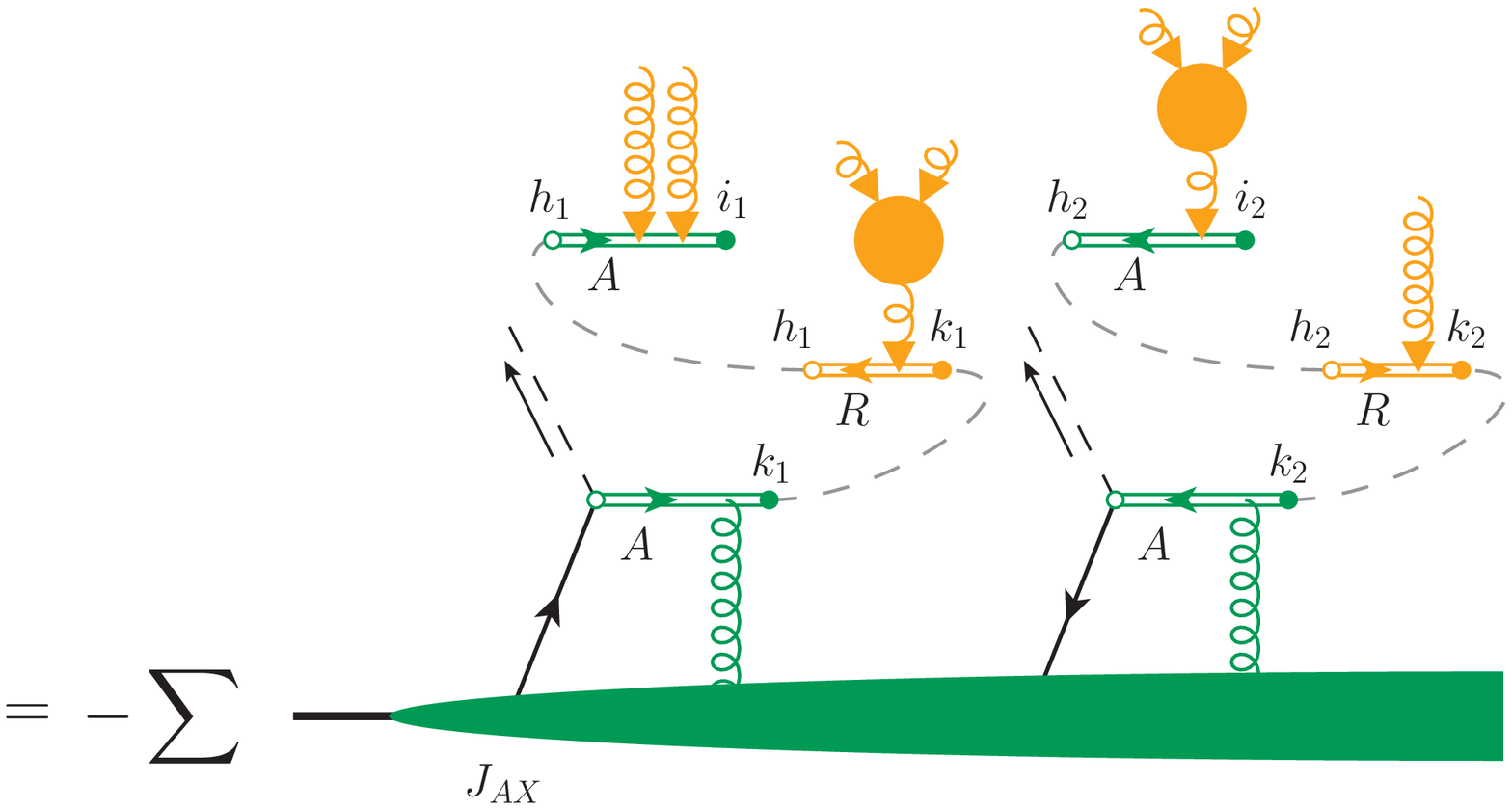} \\[1.5em]
\includegraphics[scale=0.5, trim=0 0 460 0, clip]{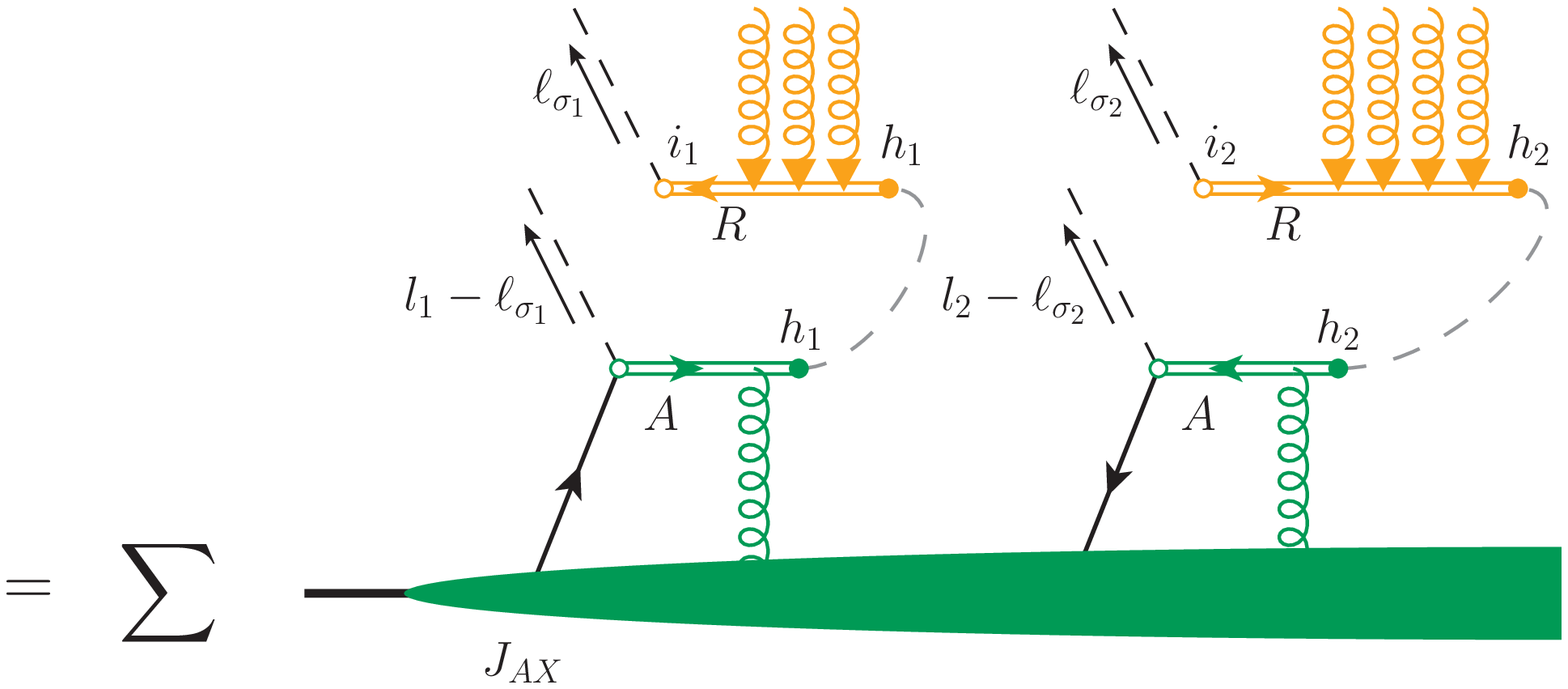}
\caption{Decoupling of $n$ soft gluons from a collinear graph in the amplitude of a DPS process.  The sums over different possibilities in the last three steps are specified in the text.}
   \label{fig:multi_gluon_attachment_dpd}
\end{figure}

The l.h.s.\ of \eqref{eqn:multi_gluon_factorised} is represented in the first row of figure~\ref{fig:multi_gluon_attachment_dpd}, and the r.h.s.\ of the equation in the last row of the figure.  We give a proof by induction with $n=1$ as base case.  It proceeds as follows.
\begin{enumerate}
\item Apply the Ward identity \eqref{JAX-ward} to the graph with $n$ external soft gluons.

An example of the graphs contributing to the r.h.s.\ of \eqref{JAX-ward} is shown in the second row of figure~\ref{fig:multi_gluon_attachment_dpd}.  The soft gluons couple either to internal collinear lines of the graph or to one of the two eikonal lines along $v_A$.  The coupling can be either direct or involve the merging of $m \to 1$ gluons in tree-level vertices $V^{(m)}$.  Only the case in which all $n$ gluons couple directly to the internal collinear lines is excluded from the sum, since this appears on the l.h.s.\ of the Ward identity.

To characterise the different graphs, we introduce a partition of the $n$ soft gluons into three sets, $\sigma$, $\tau_1$ and $\tau_2$, such that the gluons in set $\sigma$ couple to internal collinear lines and those in $\tau_i$ couple to the eikonal line associated with the first or the second parton, respectively.

\item For each of the diagrams obtained, apply the approximations discussed in section~\ref{sec:approximations}.

This implies neglecting graphs in which soft gluons couple in the middle of an eikonal line along $v_A$.  It also implies factoring the collinear and soft parts of the eikonal lines for those graphs in which all soft gluons couple at the end of a line, see \eqref{eqn:soft_approximation_wilson_lines} and \eqref{eqn:soft_approximation_adj_WL}.  Furthermore, as specified in \eqref{GY-insert}, a Grammer-Yennie factor $Y_R$ is inserted for the outgoing gluon for each of the vertices $V^{(m)}$.  We then have
\begin{align}
\label{eqn:soft_approximation_multi_gluons}
& Y_{R}^{\mu_1 \nu_1} (\ell_1) \, \dots \, Y_{R}^{\mu_n \nu_n} (\ell_n) \,
\Bigl[ J_{A X}^{i_1 i_2} \bigl( l_1, l_2; \{ \ell_n \} \bigr)
   \Bigr]^{b_1 \dots b_n}_{\nu_1 \dots \nu_n}
\nn \\
& \qquad
= {}- Y_{R}^{\mu_1 \nu_1} (\ell_1) \, \dots \, Y_{R}^{\mu_n \nu_n} (\ell_n) \,
  \sum_{ \pi, \sigma, \tau_1, \tau_2 \atop
        |\tau_1| + |\tau_2| > 0 \rule{0pt}{1.2ex} }
  \Bigl[ {\mathcal{E}}^{(|\tau_1|)}_{A, i_1 h_1}
     \bigl( \pi \{ \ell \}_{\tau_1} \bigr)
  \Bigr]_{ \pi \{ \nu \}_{\tau_1}}^{ \pi \{ b \}_{\tau_1}} \;
\nn \\[0.1em]
& \qquad\qquad \times
  \Bigl[ J_{A X}^{h_1 h_2} \bigl( l_1 - \ell_{\tau_1}, l_2 - \ell_{\tau_2};
         \pi \{ \ell \}_{\sigma} \bigr)
  \Bigr]_{ \pi \{ \nu \}_{\sigma}}^{ \pi \{ b \}_{\sigma}} \,
  \Bigl[ {\bar{\mathcal{E}}}^{(|\tau_2|)}_{A, h_2 i_2}
     \bigl( \pi \{ \ell \}_{\tau_2} \bigr)
  \Bigr]_{ \pi \{ \nu \}_{\tau_2}}^{ \pi \{ b \}_{\tau_2}}
\nn \\[0.1em]
& \qquad\quad + \{ \text{terms with vertices } V^{(m)} \} \,.
\end{align}
The terms with vertices $V^{(m)}$ are obtained from those explicitly given by selecting subsets $\ell_i \ldots \ell_j$ of gluon momenta in each of the partitions $\sigma$, $\tau_1$ and $\tau_2$ and replacing the associated Grammer-Yennie factors as
\begin{align}
Y_R^{\mu_i \nu_i} \dots Y_R^{\mu_j \nu_j}
& \to Y_R^{\mu_i \nu_i} \dots Y_R^{\mu_j \nu_j} \;
\Bigl[ V^{(m)} (\ell_i \ldots \ell_j)
  \Bigr]_{\nu_i \ldots \nu_j \mu}^{b_i \ldots b_j b} \; Y_R^{\mu \nu} \,,
\end{align}
where $m$ is the number of incoming gluons at the soft vertex, and $\nu$ and $b$ are the indices for the outgoing gluon.  The eikonal factors and $J_{A X}$ then depend on fewer gluon momenta than in the term written out in \eqref{eqn:soft_approximation_multi_gluons}, and the sum includes the case in which $|\tau_1| = |\tau_2| = 0$.

\item Use the decoupling formula \eqref{eqn:multi_gluon_factorised} with $n$ replaced by $n' < n$.

Since in all terms on the r.h.s.\ of \eqref{eqn:soft_approximation_multi_gluons} the factor $J_{A X}$ has less than $n$ incoming soft gluons, this step removes all external soft gluons from the collinear factor, and we have
\begin{align}
\label{eqn:multi_gluon_attachment_all_eikonal}
\protect\eqref{eqn:soft_approximation_multi_gluons}
&= {}- Y_{R}^{\mu_1 \nu_1} (\ell_1) \, \dots \, Y_{R}^{\mu_n \nu_n} (\ell_n) \,
  \sum_{ \pi, \sigma_1, \sigma_2, \tau_1, \tau_2 \atop
        |\tau_1| + |\tau_2| > 0 \rule{0pt}{1.2ex} }
  \Bigl[ {\mathcal{E}}^{(|\tau_1|)}_{A, i_1 h_1}
     \bigl( \pi \{ \ell \}_{\tau_1} \bigr)
  \Bigr]_{ \pi \{ \nu \}_{\tau_1}}^{ \pi \{ b \}_{\tau_1}} \;
\nn \\[0.1em]
& \qquad \times
  \Bigl[ \bar{\mathcal{E}}^{(|\sigma_1|)}_{R, h_1 k_1}
     \bigl( \pi \{ \ell \}_{\sigma_1} \bigr)
  \Bigr]_{ \pi \{ \nu \}_{\sigma_1}}^{ \pi \{ b \}_{\sigma_1}} \;
  J_{A X}^{k_1 k_2} ( l_1 - \ell_{\sigma_1} - \ell_{\tau_1},
            l_2 - \ell_{\sigma_2} - \ell_{\tau_2})
\nn \\[0.4em]
& \qquad \times
  \Bigl[ {\mathcal{E}}^{(|\sigma_2|)}_{R, k_2 h_2}
     \bigl( \pi \{ \ell \}_{\sigma_2} \bigr)
  \Bigr]_{ \pi \{ \nu \}_{\sigma_2}}^{ \pi \{ b \}_{\sigma_2}} \;
  \Bigl[ \bar{\mathcal{E}}^{(|\tau_2|)}_{A, h_2 i_2}
     \bigl( \pi \{ \ell \}_{\tau_2} \bigr)
  \Bigr]_{ \pi \{ \nu \}_{\tau_2}}^{ \pi \{ b \}_{\tau_2}}
\nn \\[0.5em]
& \quad
  + \{ \text{terms with vertices } V^{(m)} \} \,.
\end{align}
Here the set $\sigma$ of soft momenta in \eqref{eqn:soft_approximation_multi_gluons} has been partitioned into sets $\sigma_1$ and $\sigma_2$.  The terms with vertices $V^{(m)}$ are obtained in the same way as before, and again the corresponding sum includes the case $|\tau_1| = |\tau_2| = 0$.

Note that in \eqref{eqn:multi_gluon_attachment_all_eikonal} we put a Grammer-Yennie factor for each external gluon, including the gluons for which we used the decoupling formula \eqref{eqn:multi_gluon_factorised}.  This is possible because $\mathcal{E}_R$ and $\bar{\mathcal{E}}_R$ remain unchanged when their Lorentz indices are contracted with $Y_R$, as is readily seen from the definitions \eqref  {eqn:eikonal_order_n} and \eqref{eqn:adj_eikonal_order_n}.

\item Apply a Ward identity to the combination of eikonal lines.

For given values $l_1 - \ell_{\sigma_1} - \ell_{\tau_1}$ and $l_2 - \ell_{\sigma_2} - \ell_{\tau_2}$ of the momentum arguments of $J_{A X}$, the expression \eqref{eqn:multi_gluon_attachment_all_eikonal} involves a sum over the products $\mathcal{E}_{A} \, \bar{\mathcal{E}}_{R}$ and $\mathcal{E}_{R} \, \bar{\mathcal{E}}_{A}$  of eikonal lines, to which the incoming soft gluons attach either directly or via vertices $V^{(m)}$.
To simplify this sum we can use the Ward identity
\begin{align}
\label{full-ward-WL}
   k_1^{\mu_1} \dots k_{p}^{\mu_p} \;
   \Bra{ 0 \rule{0pt}{1.05em} \, }
   T A_{\mu_1}^{a_1} (k_1) \dots A_{\mu_p}^{a_p} (k_{p}) \;
   \bigl[ W_{\! A, i h}^{} \; W^\dagger_{R, h k} \bigl] (k)
   \Ket{ \, \rule{0pt}{1.05em} 0 }
 & = 0 \,,
\end{align}
where $[ W_{\! A}^{} \, W^\dagger_{R} \ms](k)$ is the Fourier transform \eqref{fourier-def} of $W_{\! A}^{}(x) \, W^\dagger_{R}(x)$.  We will prove this identity in the appendix.  Expanding the product of Wilson lines to order $g^{\ms p}$, one gets all tree level graphs in which the gluons couple to the product $\mathcal{E}_{A} \, \bar{\mathcal{E}}_{R}$ of eikonal lines either directly or via vertices $V^{(m)}$.  A pictorial representation of this result is given in figure~\ref{fig:ward_id_wilson_lines}.  An analogous identity holds for the product $[ W^{}_{R} \; W_{\! A}^{\dagger} \ms](k)$ and, truncated to the appropriate order in $g$, gives a sum over the corresponding gluon attachments to $\mathcal{E}_{R} \, \bar{\mathcal{E}}_{A}$.  It is straightforward to check \eqref{full-ward-WL} at order $g^{\ms p}$ for small values of $p$ by explicit evaluation of the Feynman graphs.

\begin{figure}
   \centering
   \includegraphics[scale=0.5]{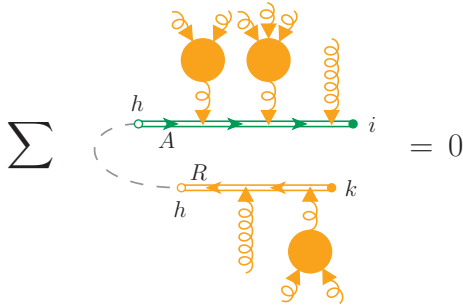}
   \caption{Pictorial representation of the Ward identity \protect\eqref{full-ward-WL} for the product of two Wilson lines.  The sum runs over all possibilities to couple the gluons to the eikonal lines, either directly or via vertices $V^{(m)}$.}
   \label{fig:ward_id_wilson_lines}
\end{figure}

The only graphs missing on the r.h.s.\ of \eqref{eqn:multi_gluon_attachment_all_eikonal} are those with $|\tau_1| = |\tau_2| = 0$ and no vertex $V^{(m)}$.  In these missing graphs, all soft gluons couple directly to either $\bar{\mathcal{E}}_{R}$ or ${\mathcal{E}}_{R}$.  We thus obtain the r.h.s.\ of \eqref{eqn:multi_gluon_factorised}, which completes the proof by induction.
\end{enumerate}

%%%%%%%%%%%%%%%%%%%%%%%%%%%%%%%%%%%%%%%%%%%%

\subsection{Soft gluon subtraction in collinear matrix elements}
\label{sec:soft-gluon-sub}

As announced in section~\ref{sec:overview}, we now return to the subtraction terms in the definition of the final collinear factor $J_{A X}(l_1, l_2)$, whose unsubtracted form is
\begin{align}
\label{JAX-unsub}
& J_{A X,\, \text{unsub}}^{i_1 i_2} (l_1, l_2 )
= \Bra{ X \rule{0pt}{1.05em} }
   T \, \Gamma_1 \bigl[ W_{\! A}^{}\, q \bigr]_{i_1}(l_1) \;
      \bigl[ \bar{q} \ms W^\dagger_{\!\bs A} \ms \bigr]_{i_2}(l_2) \Gamma_2
   \Ket{ \rule{0pt}{1.05em}P } \,.
\end{align}
The following discussion builds on the one for single parton scattering in chapter 10.8.5 of~\cite{Collins:2011zzd} and expands the argument for DPS in section~2.1 of \cite{Diehl:2015bca}.

\begin{figure}
\centering
\includegraphics[scale=0.52, trim=0 0 280 0, clip]{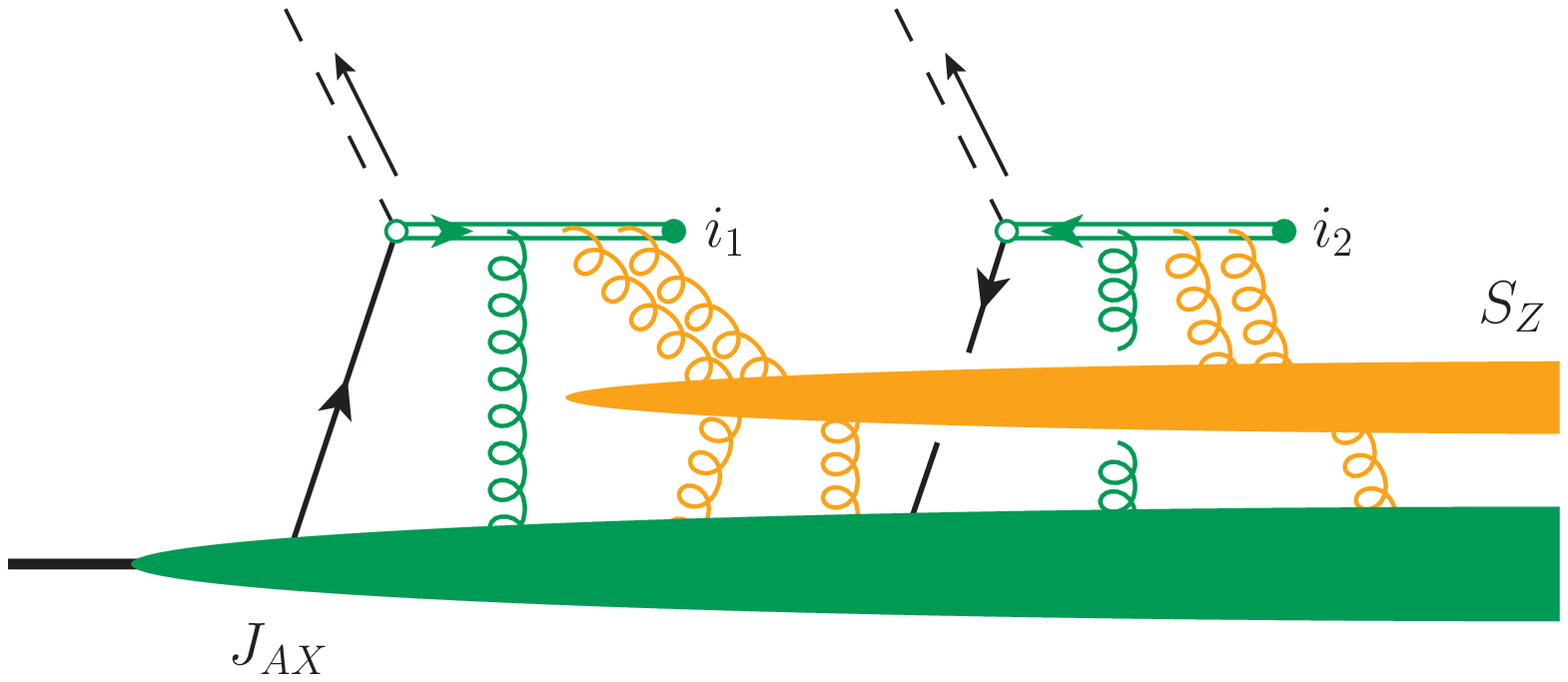} \\[1.5em]
\includegraphics[scale=0.52, trim=0 0 280 0, clip]{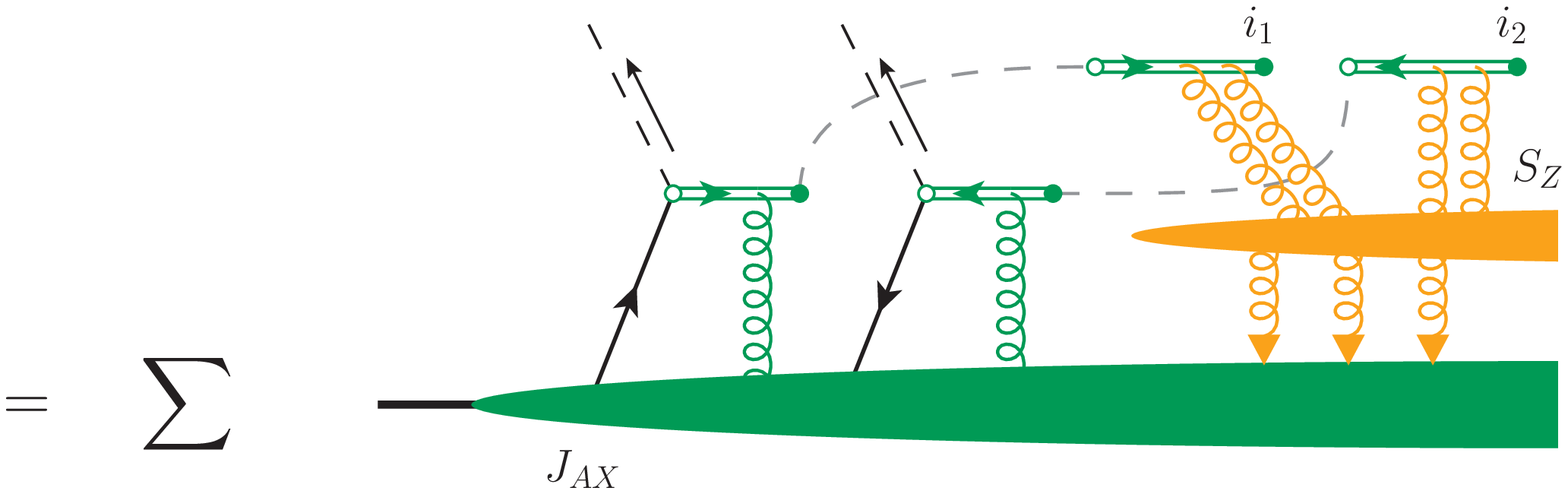} \\[1em]
\includegraphics[scale=0.52, trim=0 0 280 0, clip]{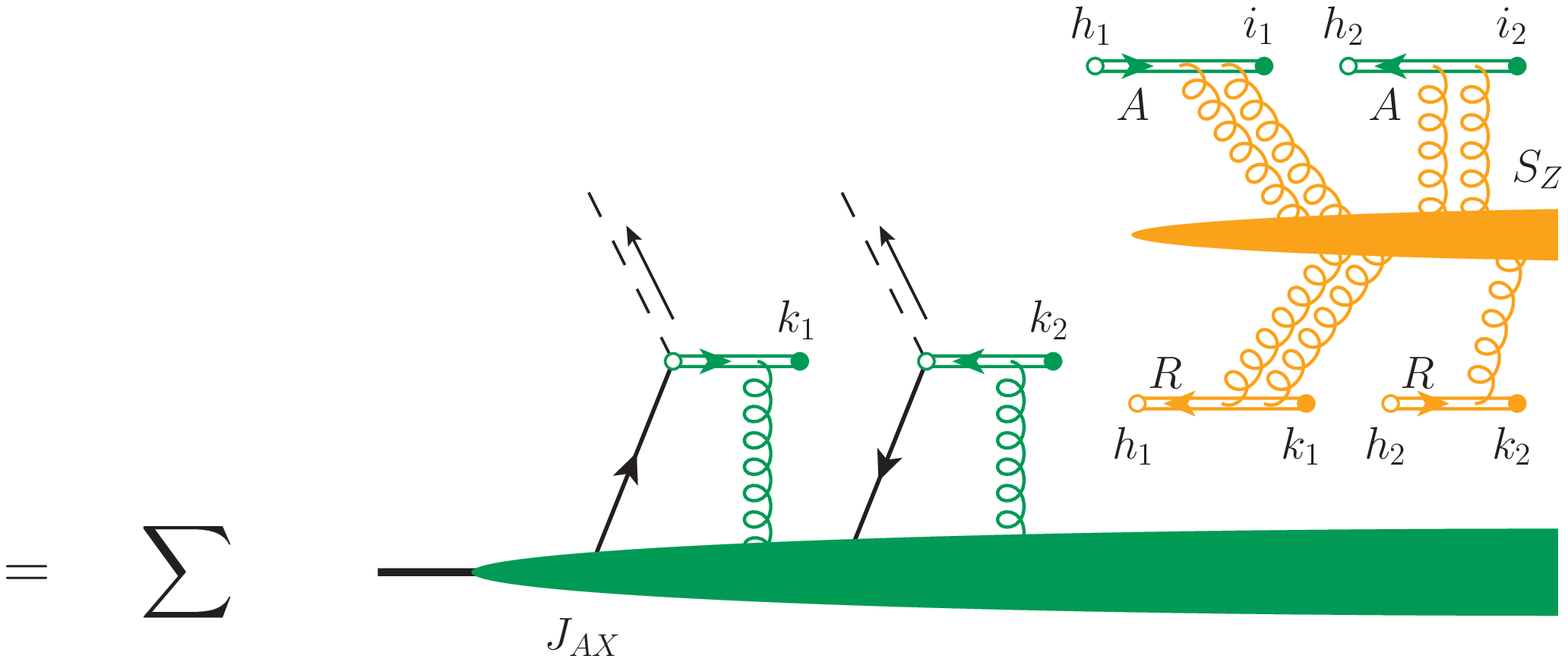}
\caption{Decoupling of soft gluons from the unsubtracted collinear matrix element $J_{A X,\, \text{unsub}}$ defined in \protect\eqref{JAX-unsub}.}
   \label{fig:JAX-unsub}
\end{figure}

The matrix element \eqref{JAX-unsub} receives a leading contribution from momentum regions in which the gluons that attach to the eikonal lines along $v_A$ are not collinear but soft.  For the same reason as in section~\ref{sec:approximations},  the soft gluons must couple at the end of an eikonal line. The unsubtracted factor $J_{A X,\, \text{unsub}}$ can hence be represented as shown in the top row of figure~\ref{fig:JAX-unsub}.  The soft gluons attached to the eikonal lines can couple among themselves in the soft subgraph $S_Z$, before coupling to the internal lines of the collinear subgraph $J_{A X}$.  Using \eqref{eqn:soft_approximation_wilson_lines} and \eqref{eqn:soft_approximation_adj_WL} one can write the eikonal factors as products of eikonal lines connected with only soft or only collinear gluons, as shown in the middle row of the figure.  One can furthermore deform the soft momenta out of the Glauber region, using the same justification as for the overall DPS cross section \cite{Diehl:2015bca}.  This allows one to insert Grammer-Yennie factors $Y_R$ for the soft gluons entering the collinear subgraph.  One can then decouple the soft gluons using the formula \eqref{eqn:multi_gluon_factorised} of the previous subsection.  The result is shown in the last row of the figure, where we have the subtracted collinear factor $J_{A X}(l_1, l_2)$ and a soft factor with eikonal lines along $v_A$ and $v_R$.

Following similar steps as those shown for the DPS amplitude in the next section, one obtains the collinear factor as a product of $J_{A X}$ and a matrix element of Wilson lines along $v_A$ and $v_R$.  Going from the amplitude to the cross section level, one finds the result given in equation~(2.13) in \cite{Diehl:2015bca}.

\section{From eikonal to Wilson lines}
\label{sec:final_result}

Having established the decoupling for an arbitrary number $n$ of soft gluons, we can now perform a sum over $n$ and convert the eikonal lines obtained in the previous section into Wilson line operators.  We do this at the level of the amplitude for the DPS cross section.

Using the decoupling formula \eqref{eqn:multi_gluon_factorised} and its analogue for $J_{B Y}$ in the contraction \eqref{ASB-GY} of soft and collinear factors and inserting the result into \eqref{lum-X-def}, we obtain
\begin{align}
\label{lumi-decoupled}
& \mathcal{L}_{X Y Z}^{i_1 i_2 \, \bar{\imath}_1 \bar{\imath}_2 \,
  (n, \bar{n})}
= \frac{1}{n! \, \bar{n}!} \,
  \int \! \fourmomdiff{l_1} \, \fourmomdiff{\bar{l}_1} \, \fourmomdiff{l_2} \, \fourmomdiff{\bar{l}_2} \;
  \prod_{k = 0}^n \, \prod_{\bar{k} = 0}^{\bar{n}}
  \int \! \fourmomdiff{\ell_k} \fourmomdiff{\bar{\ell}_{\bar{k}}}
\nn \\[0.2em]
&\quad \times
  \twopif \, \deltaf \left( q_1 - l_1 - \bar{l}_1 \right) \, \twopif \, \deltaf \left( q_2 - l_2 - \bar{l}_2 \right) \;
\nn \\[0.4em]
&\quad \times
\sum_{ \pi, \sigma_1, \sigma_2 \atop
        |\sigma_1| + |\sigma_2| = n \rule{0pt}{1.2ex} }
  \Bigl[ \bar{\mathcal{E}}^{(|\sigma_1|)}_{R, i_1 h_1}
     \bigl( \pi \bigl\{ \tilde\ell \ms\bigr\}_{\sigma_1} \bigr)
  \Bigr]_{ \pi \{ \mu \}_{\sigma_1}} \;
  J_{A X}^{h_1 h_2} \bigl(l_1 - \tilde\ell_{\sigma_1},
     l_2 - \tilde\ell_{\sigma_2} \bigr) \;
  \Bigl[ {\mathcal{E}}^{(|\sigma_2|)}_{R, h_2 i_2}
     \bigl( \pi \bigl\{ \tilde\ell \ms\bigr\}_{\sigma_2} \bigr)
  \Bigr]_{ \pi \{ \mu \}_{\sigma_2}}
\nn \\[0.2em]
&\quad \times
\sum_{ \pi, \bar\sigma_1, \bar\sigma_2 \atop
        |\bar\sigma_1| + |\bar\sigma_2| = \bar{n} \rule{0pt}{1.2ex} }
  \Bigl[ \bar{\mathcal{E}}^{(|\bar\sigma_2|)}_{L, \bar{\imath}_2 \bar{h}_2}
     \bigl( \pi \bigl\{ \tilde{\bar\ell} \ms\bigr\}_{\bar\sigma_2} \bigr)
  \Bigr]_{ \pi \{ \nu \}_{\bar\sigma_2}} \;
  J_{B Y}^{\bar{h}_1 \bar{h}_2} \bigl( \bar{l}_1
    - \tilde{\bar{\ell}}_{\bar\sigma_1},
    \bar{l}_2 - \tilde{\bar{\ell}}_{\bar\sigma_2} \bigl) \;
  \Bigl[ {\mathcal{E}}^{(|\bar\sigma_1|)}_{L, \bar{h}_1 \bar{\imath}_1}
     \bigl( \pi \bigl\{ \tilde{\bar\ell} \ms\bigr\}_{\bar\sigma_1} \bigr)
  \Bigr]_{ \pi \{ \nu \}_{\bar\sigma_1}}
\nn \\[0.4em]
&\quad \times
  \Bigl[ S_{Z}^{} \bigl( \{ \ell_n \}, \{ \bar{\ell}_{\bar{n}} \} \bigr)
  \Bigr]^{\mu_1 \ldots \mu_n, \, \nu_1 \ldots \nu_{\bar{n}}} \,.
\end{align}
For brevity we have omitted the colour indices of gluons, which are contracted between eikonal lines and the soft factor in the same way as the gluon polarisation indices.  By contrast, we write again a tilde to indicate the approximation \eqref{eqn:soft_approximation_momenta} of soft momenta in the collinear and eikonal factors.

The sum over partitions $\sigma_1, \sigma_2$ of the gluons with momenta $\ell_1 \ldots \ell_n$ and over the permutations $\pi$ of how these gluons couple to the eikonal lines can be greatly simplified.  This is because the soft factor $S_Z$ is invariant under permutations of its external gluon momenta, which is readily seen from its definition \eqref{SX-def}.  One can hence limit the sum over $\sigma_1, \sigma_2$ and $\pi$ in \eqref{lumi-decoupled} to a single partition and a single ordering.  We retain the partition in which set $n_1$ includes the gluons with momenta $\ell_1 \dots \ell_{|n_1|}$ and set $n_2$ those with momenta $\ell_{|n_1| + 1} \dots \ell_{n}$ and take the permutation in which the $\ell_k$ are ordered sequentially in $k$.  This gives a combinatorial factor
\begin{equation}
   \binom{n}{|n_1| \ms} \, |n_1| \ms ! \; |n_2| \ms ! = n! \,,
\end{equation}
where we used $|n_1| + |n_2| = n$.  The first factor on the l.h.s.\ is the number of partitions $\sigma_1, \sigma_2$, whilst the second and third factor give the number of permutations within each set.  Repeating the same argument for the gluons with momenta $\bar{\ell}_1 \dots \bar{\ell}_{\bar{n}}$, we get
\begin{align}
\label{lumi-2}
& \mathcal{L}_{X Y Z}^{i_1 i_2 \, \bar{\imath}_1 \bar{\imath}_2 \,
  (n, \bar{n})}
= \int \! \fourmomdiff{l_1} \, \fourmomdiff{\bar{l}_1} \, \fourmomdiff{l_2} \,
  \fourmomdiff{\bar{l}_2} \;
  \prod_{k = 0}^n \, \prod_{\bar{k} = 0}^{\bar{n}}
  \int \! \fourmomdiff{\ell_k} \fourmomdiff{\bar{\ell}_{\bar{k}}}
\nn \\[0.2em]
&\quad \times
  \twopif \, \deltaf \left( q_1 - l_1 - \bar{l}_1 \right) \, \twopif \, \deltaf \left( q_2 - l_2 - \bar{l}_2 \right)
\nn \\[0.4em]
&\quad \times
\sum_{ |n_1| + |n_2| = n }
  \Bigl[ \bar{\mathcal{E}}^{(|n_1|)}_{R, i_1 h_1}
     \bigl( \bigl\{ \tilde\ell \,\bigr\}_{n_1} \bigr)
  \Bigr]_{ \{ \mu \}_{n_1}} \;
  J_{A X}^{h_1 h_2} \bigl(l_1 - \tilde\ell_{n_1}, l_2 - \tilde\ell_{n_2} \bigr) \;
  \Bigl[ {\mathcal{E}}^{(|n_2|)}_{R, h_2 i_2}
     \bigl( \bigl\{ \tilde\ell \,\bigr\}_{n_2} \bigr)
  \Bigr]_{ \{ \mu \}_{n_2}}
\nn \\[0.2em]
&\quad \times
\sum_{ |\bar{n}_1| + |\bar{n}_2| = \bar{n} }
  \Bigl[ \bar{\mathcal{E}}^{(|\bar{n}_2|)}_{L, \bar{\imath}_2 \bar{h}_2}
     \bigl( \bigl\{ \tilde{\bar\ell} \,\bigr\}_{\bar{n}_2} \bigr)
  \Bigr]_{ \{ \nu \}_{\bar{n}_2}} \;
  J_{B Y}^{\bar{h}_1 \bar{h}_2} \bigl( \bar{l}_1 - \tilde{\bar{\ell}}_{\bar{n}_1},
    \bar{l}_2 - \tilde{\bar{\ell}}_{\bar{n}_2} \bigr) \;
  \Bigl[ {\mathcal{E}}^{(|\bar{n}_1|)}_{L, \bar{h}_1 \bar{\imath}_1}
     \bigl( \bigl\{ \tilde{\bar\ell} \,\bigr\}_{\bar{n}_1} \bigr)
  \Bigr]_{ \{ \nu \}_{\bar{n}_1}}
\nn \\[0.4em]
&\quad \times
\Braket{ Z | T A^{\mu_1} (\ell_1) \dots A^{\mu_n} (\ell_n) \,
   A^{\nu_1} (\bar{\ell}_1) \dots
     A^{\nu_{\bar{n}}} (\bar{\ell}_{\bar{n}})| 0 }  \,,
\end{align}
where we have explicitly written out the soft factor using \eqref{SX-def}.  We now shift the integration momenta such that $J_{A X}$ and $J_{B Y}$ become independent of soft gluon momenta:
\begin{align}
\label{lumi-3}
& \mathcal{L}_{X Y Z}^{i_1 i_2 \, \bar{\imath}_1 \bar{\imath}_2 \,
  (n, \bar{n})}
= \int \! \fourmomdiff{l_1} \, \fourmomdiff{\bar{l}_1} \, \fourmomdiff{l_2} \,
  \fourmomdiff{\bar{l}_2} \;
  \prod_{k = 0}^n \, \prod_{\bar{k} = 0}^{\bar{n}}
  \int \! \fourmomdiff{\ell_k} \fourmomdiff{\bar{\ell}_{\bar{k}}}
\nn \\
&\quad \times
  \sum_{ |n_1| + |n_2| = n  \atop
     |\bar{n}_1| + |\bar{n}_2| = \bar{n} \rule{0pt}{1.2ex} }
  \twopif \, \deltaf \bigl( q_1 - l_1 - \bar{l}_1
             - \tilde\ell_{n_1} - \tilde{\bar{\ell}}_{\bar{n}_1} \bigr) \,
  \twopif \, \deltaf \bigl( q_2 - l_2 - \bar{l}_2
             - \tilde\ell_{n_2} - \tilde{\bar{\ell}}_{\bar{n}_2} \bigr)
\nn \\[0.1em]
&\quad \times
  \Bigl[ \bar{\mathcal{E}}^{(|n_1|)}_{R, i_1 h_1}
     \bigl( \bigl\{ \tilde\ell \,\bigr\}_{n_1} \bigr)
  \Bigr]_{ \{ \mu \}_{n_1}} \;
  J_{A X}^{h_1 h_2} (l_1, l_2) \;
  \Bigl[ {\mathcal{E}}^{(|n_2|)}_{R, h_2 i_2}
     \bigl( \bigl\{ \tilde\ell \,\bigr\}_{n_2} \bigr)
  \Bigr]_{ \{ \mu \}_{n_2}}
\nn \\[0.4em]
&\quad \times
  \Bigl[ \bar{\mathcal{E}}^{(|\bar{n}_2)|}_{L, \bar{\imath}_2 \bar{h}_2}
     \bigl( \bigl\{ \tilde{\bar\ell} \,\bigr\}_{\bar{n}_2} \bigr)
  \Bigr]_{ \{ \nu \}_{\bar{n}_2}} \;
  J_{B Y}^{\bar{h}_1 \bar{h}_2} (\bar{l}_1, \bar{l}_2) \;
  \Bigl[ {\mathcal{E}}^{(|\bar{n}_1)|}_{L, \bar{h}_1 \bar{\imath}_1}
     \bigl( \bigl\{ \tilde{\bar\ell} \,\bigr\}_{\bar{n}_1} \bigr)
  \Bigr]_{ \{ \nu \}_{\bar{n}_1}}
\nn \\[0.4em]
&\quad \times
\Braket{ Z | T A^{\mu_1} (\ell_1) \dots A^{\mu_n} (\ell_n) \,
   A^{\nu_1} (\bar{\ell}_1) \dots
     A^{\nu_{\bar{n}}} (\bar{\ell}_{\bar{n}})| 0 } \,.
\end{align}
Recalling that $\ell_{n_1}$ and $\ell_{n_2}$ are defined as sums over soft momenta, we can insert explicit integrations
\begin{align}
& \int \fourmomdiff{\ell_{n_1}} \;
  \twopif \, \deltaf \biggl( \ell_{n_1} - \sum_{k=1}^{n_1} \ell_k \biggr) \;
  \int \fourmomdiff{\ell_{n_2}} \;
  \twopif \, \deltaf \biggl( \ell_{n_2} - \sum_{k=n_1 + 1}^{n} \ell_k \biggr)
\end{align}
as well as their analogues for $\bar{\ell}_{\bar{n}_1}$ and $\bar{\ell}_{\bar{n}_2}$.
Combining the delta functions in these expressions with the integrations over $\ell_k$ and $\bar{\ell}_{\bar{k}}$, the gluon field operators in the soft factor, and the eikonal lines in \eqref{lumi-3}, we recognise the expressions \eqref{WL-FT} and \eqref{WL-adj-FT} of Wilson lines in momentum space (expanded to the appropriate order in $g$).  We thus obtain
\begin{align}
\label{lumi-4}
& \mathcal{L}_{X Y Z}^{i_1 i_2 \, \bar{\imath}_1 \bar{\imath}_2 \,
  (n, \bar{n})}
= \int \! \fourmomdiff{l_1} \, \fourmomdiff{\bar{l}_1} \,
  \fourmomdiff{l_2} \, \fourmomdiff{\bar{l}_2} \,
  \sum_{ n_1 + n_2 = n  \atop \bar{n}_1 + \bar{n}_2 = \bar{n} \rule{0pt}{1.2ex} }
  \int \! \fourmomdiff{\ell_{n_1}} \, \fourmomdiff{\ell_{n_2}} \,
  \fourmomdiff{\bar{\ell}_{\bar{n}_1}} \, \fourmomdiff{\bar{\ell}_{\bar{n}_2}}
\nn \\
&\quad \times
   \twopif \, \deltaf \bigl( q_1 - l_1 - \bar{l}_1 - \tilde\ell_{n_1}
     - \tilde{\bar{\ell}}_{\bar{n}_1} \bigr) \,
   \twopif \, \deltaf \bigl( q_2 - l_2 - \bar{l}_2 - \tilde\ell_{n_2}
     - \tilde{\bar{\ell}}_{\bar{n}_2} \bigr)
\nn \\[0.4em]
&\quad \times
  \int \! \ddf{\zeta_1} \, \ddf{\zeta_2} \,
  \ddf{\bar{\zeta}_1} \, \ddf{\bar{\zeta}_2} \;
  \exp\bigl[ i ( \ell_{n_1} \cdot \zeta_1^{} + \ell_{n_2} \cdot \zeta_2^{}
    + \bar{\ell}_{\bar{n}_1} \cdot \bar{\zeta}_1^{}
    + \bar{\ell}_{\bar{n}_2} \cdot \bar{\zeta}_2^{}) \bigr]
\nn \\
&\quad \times
\Braket{ Z | \Bigl[ W_R^{\dagger (n_1)} (\zeta_1) \Bigr]_{i_1 h_1} \,
  \Bigl[ W_R^{(n_2)} (\zeta_2) \Bigr]_{h_2 i_2} \,
  \Bigl[ W_L^{\dagger (\bar{n}_1)} (\bar{\zeta}_2)
    \Bigr]_{\bar{\imath}_2 \bar{h}_2} \,\
  \Bigl[ W_L^{(\bar{n}_2)} (\bar{\zeta}_1)
    \Bigr]_{\bar{h}_1 \bar{\imath}_1} | 0 }
\nn \\[0.4em]
&\quad \times
   J_{A X}^{h_1 h_2} (l_1, l_2) \, J_{B Y}^{\bar{h}_1 \bar{h}_2}
   (\bar{l}_1, \bar{l}_2) \,,
\end{align}
where we have explicitly written out the Fourier transform of the Wilson lines.  Instead of sets of gluon momenta, $n_1$ and $n_2$ are now integers (equal to the size of these sets).  The delta functions in the second line of \eqref{lumi-4} can be simplified by neglecting the small plus or minus components of soft momenta compared with the large components of $q_1, q_2$ and the collinear momenta.  The transverse components of soft momenta must however be kept.  Representing the corresponding part of the delta functions as  Fourier integrals, we have
\begin{align}
\label{delta-approx}
\delta^{(D)} \bigl( q - l - \bar{l} - \tilde\ell - \tilde{\bar{\ell}} \,\bigr)
  & \approx \delta ( q^+ - l^+ ) \, \delta ( q^- - \bar{l}^- )
   \int\! \frac{\dd^{D-2} \tvec{\xi}}{(2\pi)^{D-2}} \; e^{ - i
   ( \boldsymbol{q} - \boldsymbol{l} - \boldsymbol{\bar{l}}
   - \boldsymbol{\ell} - \boldsymbol{\bar{\ell}} ) \cdot \tvec{\xi} } \,,
\end{align}
where we omitted indices like $1$ or $n_1$ for brevity.  Inserting this into \eqref{lumi-4}, we can perform the integrals over the soft momenta, the positions of the Wilson line operators, and over the large plus or minus components constrained by \eqref{delta-approx}.  This gives
\begin{align}
\label{lumi-5}
& \mathcal{L}_{X Y Z}^{i_1 i_2 \, \bar{\imath}_1 \bar{\imath}_2 \,
  (n, \bar{n})}
= \int\! \dd^{D-2} \tvec{\xi}_1 \, \dd^{D-2} \tvec{\xi}_2 \;
  e^{-i \tvec{q}_1 \cdot \tvec{\xi}_1 - i \tvec{q}_2 \cdot \tvec{\xi}_2}
  \nn \\[0.2em]
& \quad \times
  \sum_{ n_1 + n_2 = n  \atop \bar{n}_1 + \bar{n}_2 = \bar{n} \rule{0pt}{1.2ex} }
  \Braket{ Z | \Bigl[ W_R^{\dagger (n_1)} (\tvec{\xi}_1) \Bigr]_{i_1 h_1} \,
  \Bigl[ W_R^{(n_2)} (\tvec{\xi}_2) \Bigr]_{h_2 i_2} \,
  \Bigl[ W_L^{\dagger (\bar{n}_1)} ({\tvec{\xi}}_2)
    \Bigr]_{\bar{\imath}_2 \bar{h}_2} \,\
  \Bigl[ W_L^{(\bar{n}_2)} ({\tvec{\xi}}_1)
    \Bigr]_{\bar{h}_1 \bar{\imath}_1} | 0 }
\nn \\
& \quad \times
  \int\! \frac{\dd l_1^-\, \dd^{D-2} \tvec{l}_1^{}}{(2\pi)^{D-1}} \,
  \frac{\dd l_2^-\, \dd^{D-2} \tvec{l}_2^{}}{(2\pi)^{D-1}} \;
  e^{i \tvec{l}_1^{} \cdot \tvec{\xi}_1 + i \tvec{l}_2^{} \cdot \tvec{\xi}_2} \;
  J_{A X}^{h_1 h_2} (l_1, l_2) \,\Big|_{l_1^+ = q_1^+,\, l_2^+ = q_2^+}
\nn \\[0.4em]
& \quad \times
  \int\! \frac{\dd \bar{l}_1^-\, \dd^{D-2} \bar{\tvec{l}}_1^{}}{(2\pi)^{D-1}} \,
  \frac{\dd \bar{l}_2^-\, \dd^{D-2} \bar{\tvec{l}}_2^{}}{(2\pi)^{D-1}} \;
  e^{i \bar{\tvec{l}}_1^{} \cdot \tvec{\xi}_1
   + i \bar{\tvec{l}}_2^{} \cdot \tvec{\xi}_2} \;
  J_{B Y}^{\bar{h}_1 \bar{h}_2} (\bar{l}_1, \bar{l}_2)
  \,\Big|_{\bar{l}_1^- = q_1^-,\, \bar{l}_2^+ = q_2^-} \,,
\end{align}
where the boldface arguments of Wilson lines indicate that their plus and minus positions are zero.  We note that the collinear factors in the last two lines are Fourier transformed to the same transverse positions as the Wilson lines with which they are connected by colour indices.  Likewise, the transverse positions of partons $1$ and $2$ in the right and left moving proton coincide, in agreement with the physical picture that partons that enter in a hard subprocess must be at the same position in the transverse plane.

Summing \eqref{lumi-5} over all $n$ and $\bar{n}$, one recovers full Wilson lines.  Combining this with the corresponding expression for the complex conjugate amplitude of the DPS process and summing over the intermediate states $X, Y$ and $Z$, one obtains unsubtracted collinear factors $J_A$ and $J_B$, as well as a soft factor $S$ defined as the vacuum expectation value of Wilson line operators.  This is the starting point of the considerations in \cite{Buffing:2017mqm}, where the colour structure of these factors is analysed, and where they are combined into double parton distributions that appear in the final cross section formula.

We shall not repeat this analysis here, but note only that the colour indices in \eqref{lumi-5} are eventually contracted with $\delta_{i_1 \bar{\imath}_1}$ and $\delta_{i_2 \ms \bar{\imath}_2}$.  For the production of colourless particles in TMD factorisation, this simply follows from the colour structure of the hard scattering subgraphs.  For collinear factorisation, where hard unobserved partons can be radiated into the final state, this is the result of nontrivial simplifications (see section~4.5 in \cite{Buffing:2017mqm}).

\section{Conclusions}
\label{sec:sum}

In this paper, we have filled a gap in the proof of factorisation for DPS processes by showing that -- after deformation out of the Glauber region -- soft gluons can be decoupled from the right or left collinear subgraphs associated with the incoming protons.  It turns out that the complicated colour structure of the collinear subgraphs, with two partons emerging in the amplitude and two in the complex conjugate one, does not pose particular problems but merely complicates the combinatorics of the proof compared with the case of SPS.  On the technical side, the  approximation \eqref{eqn:soft_approximation_momenta} of soft gluon momenta entering collinear graphs, which is weaker than in the work of Collins, Soper and Sterman, can be handled as shown in section~\ref{sec:approximations}.  This requires neglecting power suppressed terms, which is necessary in other parts of the proof as well (in particular, for dropping graphs in which soft gluons couple in the middle of an eikonal line).

Note that the main part of our argument, presented in section~\ref{sec:proof}, does not make use of the colour structure of the hard scattering subprocesses and is valid for the production of colourless particles as well as coloured ones.  However, other parts of the factorisation proof, in particular the discussion of the Glauber region, are more complicated if coloured particles are produced.  We recall that for this reason TMD factorisation has only been established for the production of colourless particles, even in SPS \cite{Rogers:2010dm,Boer:2017hqr}.

It is straightforward to extend our proof from DPS to the case of $N$ hard scatters.  The result involves a soft factor with more Wilson line operators than for $N=2$, and the number of colour channels for the collinear and soft factors is higher.  This complicates the analysis but does not pose any conceptual problems.

There are some technical issues that merit further investigation, in particular the missing proof of Ward identities for graphs with eikonal lines and fixed internal momenta, as discussed at the beginning of section~\ref{sec:proof}.  Further open problems are described in section~9 of \cite{Diehl:2017wew} and in section~4.1 of \cite{Boer:2017hqr}.  Nevertheless, we think that the present work contributes to putting factorisation formulae for DPS processes on theoretically firm ground.

\appendix

\section{Nonabelian Ward identities}
\label{app:ward}

The purpose of this appendix is to derive the Ward identities \eqref{full-ward} and \eqref{full-ward-WL} for correlation functions involving Wilson lines.  Our treatment is based on the textbook proofs given for instance in \cite{Sterman:1994ce,Itzykson:1980rh}, which use path integral methods and apply to full amplitudes.  As already discussed, the proof of soft gluon decoupling requires \eqref{full-ward} in a more restricted setting with fixed internal momenta in Feynman graphs.  A corresponding proof should be possible using the diagrammatic approach in  \cite{tHooft:1971akt,tHooft:1972qbu}, but this is beyond the scope of this paper.

Consider a theory with generic fields $\phi_i$, a Lagrangian density $\mathcal{L}(\phi_i)$, and an infinitesimal transformation $\phi_i \to \phi_i + \delta\phi_i(\{\phi_j\})$ of the fields. The corresponding Schwinger-Dyson equation for a generic $n$ point correlation function is given by
\begin{align} \label{schwingerdyson}
\delta \bra{0}  T  \phi_1(x_1) \dots \phi_n(x_n) \ket{0}
 &= \sum_{i=1}^n \Braket{0 | T  \phi_1(x_1) \dots
    \delta \phi_i\bigl( \{\phi_j(x_j)\} \bigr) \dots \phi_n(x_n) | 0}
\nn \\
 &= -i \int \! \ddf y \, \Braket{0 | T \phi_1(x_1) \dots \phi_n(x_n) \; \delta \mathcal{L}\bigl( \{\phi_i(y)\} \bigr) | 0} \,,
\end{align}
where in the first line we applied the field transformation to each field of the $n$ point function. Note that the correlation function is a vacuum expectation value.

We apply \eqref{schwingerdyson} to QCD amplitudes and take $\delta$ to be the infinitesimal BRST transformation.  This leaves the Lagrangian invariant, so that the r.h.s.\ of \eqref{schwingerdyson} is zero.  The transformation acts on the elementary fields in the Lagrangian as
\begin{align} \label{eqn:brst_transformation}
   \delta \ms q_i (x) &= i g \, t^a_{i j} \, c^a (x) \, \eta \, q_j(x) \,,
\nn \\[0.1em]
   \delta \ms \bar{q}_i (x) &= - i g \, \bar{q}_j(x) \, t^a_{j i} \, c^a (x) \, \eta  \,,
\nn \\[0.1em]
   \delta \ms A^a_\mu (x) &= \bigl[ \partial_\mu + g f^{a b c} A^c_\mu(x) \bigr] \, c^b(x) \, \eta \,,
\nn \\
   \delta \ms c^a (x) &= -\frac{1}{2}\, g\ms f^{a b c} \, c^b(x) \, c^c(x) \, \eta, \nn \\
   \delta \ms \bar{c}^a(x) &= \partial_\mu A^{a,\mu}(x)\, \eta \,,
\end{align}
where $\eta$ is an infinitesimal Grassmann parameter and Feynman gauge is assumed.  As a consequence of \eqref{eqn:brst_transformation} the gluon field strength transforms as
\begin{equation}
\delta \ms G^a_{\mu\nu} (x)
   = g \ms f^{a b c} \, c^b(x) \, \eta \, G^c_{\mu\nu} (x) \,.
\end{equation}
The fields $\phi_i$ in \eqref{schwingerdyson} are not restricted to be elementary fields, but they can be nonlocal products of elementary fields.  Indeed, the correlation functions we consider involve nonlocal operators of the type $W_{i j} (x) \ms q_j(x)$ or $W^{ab} (x) \ms G^b(x)$.

As a generalisation of \eqref{schwingerdyson}, we need identities for operator matrix elements with an external proton state $\ket{P}$ and partonic final states $\bra{X}$.  These can be rewritten as vacuum expectation values using the LSZ reduction formula.  One finds that the BRST variation of the fields associated with on-shell external particles does not contribute to the sum over $i$ in \eqref{schwingerdyson}.  For the proton this is because its interpolating field is gauge invariant and hence also BRST invariant.  For the on-shell parton states in $\bra{X}$ it follows from the truncation of propagators in the reduction formula (see e.g.\ chapter 11.2 of \cite{Sterman:1994ce} or chapter 7.4 of \cite{Peskin:1995ev}) and from the restriction to physical gluon polarisation.

The desired Ward identity \eqref{full-ward} can be obtained from
\begin{align}
\label{eqn:ward_identity_n_brst}
0 &= \left( \der{}{\eta}\, \delta \right)^n \,
  \Braket{X| T \, \bar{c} (z_1) \dots \bar{c} (z_n) \, \phi_1(x_1) \, \phi_2(x_2)  | P}
\nn \\
& = \Braket{X| T \, \partial^{\mu_1} \! {A}_{\mu_1} (z_1) \dots
    \partial^{\mu_n} \! {A}_{\mu_n} (z_n) \, \phi_1(x_1) \, \phi_2(x_2)| P}
  + \{ \text{terms with $\delta \phi_1$ or $\delta \phi_2$} \} \,,
\end{align}
where
\begin{align}
\phi_1(x) &= W_{\! A}^{}(x)\, q(x) \,,
&
\phi_2(x) &= \bar{q}(x) \ms W^\dagger_{A}(x) \,,
\end{align}
and where for brevity we have omitted colour indices.  To obtain \eqref{eqn:ward_identity_n_brst}, we applied \eqref{schwingerdyson} $n$ times, removing the Grassmann parameter $\eta$ by differentiation after each step.  Note that applying $(\dd / \dd \eta)\, \delta$ more than once to any of the fields on the l.h.s.\ of \eqref{eqn:brst_transformation} gives zero.

We obtain \eqref{full-ward} by Fourier transform to momentum space if we can show that the terms with $\delta \phi_1$ or $\delta \phi_2$ on the r.h.s.\ of \eqref{eqn:ward_identity_n_brst} are zero.  These terms contain at least one of the BRST variations $\delta \phi_1$ and $\delta \phi_2$.  To proceed, we make the limit of infinite path length in $W_{\! A}$ explicit by writing
\begin{align} \label{eqn:finite_wilson_line_def}
W_{A, i j} (x) &= \lim_{t \to -\infty} W_{A, i j} (y_t,x)
\end{align}
where $y_t = x + t \ms v_{\bs A}$ is the end of the path in
\begin{align}
W_{A, i j} (y_t,x) &= \mathcal{P} \exp \left[ \ms i g \! \int_{t}^0 \! \dd s \,
    v_{\bs A} \cdot A^a (x + s \ms v_{\bs A}) \, t^a_{i j} \right] \,.
\end{align}
From the BRST transformation
\begin{equation} \label{eqn:brst_variation_wilson}
\delta W(y,x)
= i g \left[\ms t^b c^b(y) \, \eta \, W(y,x)
    - W(y,x) \, t^b c^b(x) \, \eta \ms\right]
\end{equation}
of a Wilson line one readily obtains
\begin{align} \label{eqn:brst_variation_dressed}
\delta \ms \bigl[ W(y, x) \, q(x) \bigr]{}_i^{}
   &= i g \, t^b_{i j} \, c^b (y) \, \eta \,
      \bigl[ W(y, x) \, q(x) \bigr]{}_j^{} \,,
\nn \\
\delta \ms \bigl[ \bar{q} (x) \, W^\dagger (y, x) \bigr]{}_i^{}
   &= -i g \bigl[ \bar{q} (x) \, W^\dagger (y, x) \bigr]{}_j^{} \;
      t^b_{j i} \, c^b (y) \, \eta \,.
\end{align}
The analogous relation for gluon fields reads
\begin{align}
\delta \ms \bigl[  W(y, x) \, G(x) \bigr]^a
   &= g f^{a b c} \, c^b (y) \, \eta \, \bigl[ W(y, x) \, G(x) \bigr]^c \,.
\end{align}
The reason why the terms with $\delta \phi_1$ or $\delta \phi_2$ do not contribute in \eqref{eqn:ward_identity_n_brst} is the limit $t\to - \infty$.  To see this, we consider the variation of $W_{\! A}(y_t,x) \, q(x)$, isolate the term of order $g$ in the Wilson line and Fourier transform to momentum space:
\begin{align}
\label{ghost-element}
&\int \! \ddf x \; e^{i l \cdot x} \,
   \Bra{X} \dots c (y)\,  W_{\! A}^{(1)}(y_t, x) \, q(x) \dots \Ket{P}
\nn \\
&\quad \propto \int \ddf x \, \int_t^0 \! \dd s \int \!
   \fourmomdiff{q} \fourmomdiff{k} \fourmomdiff{p} \; e^{i l \cdot x} \, e^{-i q \cdot (x + t v_A)} \, e^{-i k \cdot(x + s v_A)} \, e^{-i p \cdot x} \,
\nn \\[0.2em]
&\qquad \times
   \Bra{X} \dots c(q) \; v_A \!\cdot\! {A}(k) \; q(p) \dots \Ket{P}
\nn \\[0.2em]
&\quad = \int \! \fourmomdiff{q} \; e^{-i t q \cdot v_A}
   \int \! \fourmomdiff{k} \; \frac{i}{k \cdot v_A + i \epsilon} \,
   \bigl( 1 - e^{-i t k \cdot v_A} \bigr) \,
\nn \\[0.2em]
&\qquad \times
   \Bra{X} \dots {c}(q) \; v_A \!\cdot\! {A} (k) \; q (l - k - q)  \dots \Ket{P} \,,
\end{align}
where the ellipsis stands for the remaining operators in the correlation function.  For large $t$ the factor $e^{-i t q \cdot v_A}$ produces strong oscillations, whereas the remaining integrand has a smooth dependence on $q$.  This gives a vanishing integral over $q$ in the limit $t \to -\infty$.  The same holds if $W^{(1)}_{\! A}$ in \eqref{ghost-element} is replaced with $W^{(n)}_{\! A}$ for arbitrary $n$, and if one considers the operator $\phi_2$ instead of $\phi_1$.

In the same manner, we can derive the Ward identity \eqref{full-ward-WL}.  Instead of \eqref{eqn:ward_identity_n_brst} we then consider
\begin{align}
\label{ward_two_WLs}
0 &= \left( \der{}{\eta}\, \delta \right)^p \,
  \Braket{X| T \, \bar{c} (z_1) \dots \bar{c} (z_p) \, \phi(x) | P}
\nn \\
& = \Braket{X| T \, \partial^{\mu_1} \! {A}_{\mu_1} (z_1) \dots
    \partial^{\mu_p} \! {A}_{\mu_p} (z_p) \, \phi(x)| P}
  + \{ \text{terms with $\delta \phi$} \}
\end{align}
with
\begin{align}
\phi(x) &= W_{\! A}^{}(x)\, W_{R}^\dagger(x)
  = \lim_{t \to -\infty} \, \lim_{u \to -\infty}
    W_{\! A}(y_t,x)\, \bigl[ W_{R}(z_u,x) \bigr]^\dagger \,,
\end{align}
where $y_t = x + t \ms v_{\bs A}$ and $z_u = x + u \ms v_{\bs R}$.  Using \eqref{eqn:brst_variation_wilson} one finds that the BRST variation of the r.h.s.\ gives a ghost field $c(y_t)$ or $c(z_u)$.  Just as in \eqref{ghost-element} this results in vanishing loop integrals when $t$ and $u$ are taken to negative infinity.  Note that the ghost field $c(x)$ in \eqref{eqn:brst_variation_wilson}  cancels in the transformation of a product of Wilson lines that meet at the same point $x$.

%%%%%%%%%%%%%%%%%%%%%%%%%%%%%%%%%%%%%%%%%%%%%%%%%%%%%%%

\acknowledgments

We gratefully acknowledge discussions with Maarten Buffing, Johannes Michel, Dave Soper and Frank Tackmann.
Some calculations useful in the development of this work were done using FORM \cite{Ruijl:2017dtg}, and all pictures were produced using JaxoDraw \cite{Binosi:2003yf,Binosi:2008ig}.

%%%%%%%%%%%%%%%%%%%%%%%%%%%%%%%%%%%%%%%%%%%%%%%%%%%%%%%

\bibliographystyle{JHEP}

\phantomsection
\addcontentsline{toc}{section}{References}
\bibliography{ward}

\end{document}